\documentclass[fleqn,usenatbib]{mnras}
\usepackage{newtxtext,newtxmath}
\usepackage[T1]{fontenc}
\usepackage{ae,aecompl}
\usepackage{graphicx,amsmath,booktabs,lastpage,capt-of}
\hypersetup{hypertexnames=false}
\newenvironment{SURDtable}{\begin{table*}}{\end{table*}}
\newenvironment{SURDfigure}{\begin{figure*}}{\end{figure*}}

\title{Season by season causal analysis of the continuum and velocity-resolved H$\beta$ emission of NGC~5548}
\author[A. Mitra \& V. Zarikas]{
Ayan Mitra,$^{1,2}$\thanks{E-mail: ayan@illinois.edu},
Vasilios Zarikas$^{3}$\thanks{E-mail: vzarikas@uth.gr}\\
$^{1}$Center for AstroPhysical Surveys, National Center for Supercomputing Applications, Urbana, IL 61801, USA\\
$^{2}$Department of Astronomy, University of Illinois Urbana-Champaign, Urbana, IL 61801, USA\\
$^{3}$Department of Mathematics, University of Thessaly, Lamia 35100, Greece
}
\date{}
\pubyear{2026}
\begin{document}
\label{firstpage}
\pagerange{\pageref{firstpage}--\pageref{LastPage}}
\maketitle
\begin{abstract}
We investigate possible causal influences in five historical seasons (1989--1993) of NGC~5548 using an adapted for the present analysis, Synergistic--Unique--Redundant Decomposition, (SURD), method. It is an information theoretic causal discovery method for time series. Two four-variable sets contain the blue wing ($B$), core ($C$) and red wing ($R$) of H$\beta$ together with either total H$\beta$ ($H$) or the 5100~\AA\ continuum ($F$). We examine 7200 decompositions with common and unequal lags, two or three seasonal flux states, and base delays of 1--30~d. We rank the individual named components within each information family, reporting their amplitudes and simultaneous leakage. In 1989, unique continuum information about the core reaches 0.799 and 0.387 of target entropy for two and three states, with leakage 0.047 and 0.276. In 1992, the red-wing history supplies unique information about the future blue wing in both variable sets and both state counts, with peak target-entropy fractions 0.106--0.166. Distinct joint influences also emerge: for the 1991 core, the two-state $S(F,B)$ and $S(F,R)$ components contribute 0.187 and 0.125 at 20~d with leakage 0.087, although the unique continuum term vanishes. Strong core--wing redundancy identifies candidate inputs for simpler predictive models. Large late-lag leakage, reaching approximately 0.97 in the total set, identifies targets for additional driver measurements and richer temporal histories. The resulting seasonal map of individual, shared and joint influences complements the response delays and velocity structure measured by reverberation mapping, and provides specific causal hypotheses for the evolving broad-line region.
\end{abstract}

\section{Introduction}

Reverberation mapping uses the delayed response of broad emission lines to continuum variability to constrain the size and kinematics of the broad-line region (BLR) \citep{Peterson1993, Peterson2004ApJ}. The relationship between the BLR size and AGN continuum luminosity
has been established through extensive reverberation-mapping
observations \citep{Bentz2009,Bentz2013}. NGC~5548 provides a long monitoring baseline on which to examine changes in this response. Recent seasonal analyses by \citet{Lu2022} and \citet{Xi2025} demonstrate that its velocity-resolved lag profiles change with epoch. A profile's asymmetry can be informative about motion and illumination, although a kinematic interpretation also depends on the response geometry and observational definition of the line bands.

Approaches to causal discovery from time series include Granger causality
\citep{Granger1969}, transfer entropy \citep{Schreiber2000}, convergent
cross mapping \citep{Sugihara2012}, and conditional-independence methods
such as PCMCI \citep{Runge2019PCMCI}. Closely related information-theoretic
developments include partial information decomposition
\citep{WilliamsBeer2010} and its extension to distinguishing unique,
redundant and synergistic information transfer from multiple sources
\citep{WilliamsBeer2011}. SURD
\citep{MartinezSanchez2024SURD,MartinezSanchez2025Thesis} addresses a
complementary multivariate question: how is the information about a
future target distributed among individual predictors and their
combinations? It distinguishes unique information, information shared
by several predictors, and information accessible through their joint
states. It also retains an information leak, the target uncertainty
unresolved by the specified predictor vector. These distinctions matter
when a line responds to a common driving continuum and when neighbouring
velocity bands have correlated histories. A small unique term can
coexist with substantial shared or joint information.

We follow the seasonal approach of the recent campaigns using the historical AGN Watch continuum and the corrected H$\beta$ component fluxes published by \citet{WandersPeterson1996}. The total line and its components form one four-variable set; replacing the total by the optical continuum forms a second. Each member is used as a target. Our main results identify the leading named unique, redundant and synergistic components for each epoch, target and flux discretization. We pair each component with its contemporaneous leakage and compare the common- and unequal-lag configurations. This resolves which individual histories and combinations carry the observed information, including modest but distinct joint contributions in a noisy astrophysical system. The extended calculations use the archived component values and preserve their original normalization alongside a target-entropy scale. 

Section~\ref{sec:surd_method} defines the method and normalizations. Section~\ref{sec:epoch_surd_results} presents the seasonal results, retains the literature comparison in Section~\ref{sec:comparison}, and gives a separate astrophysical interpretation in Section~\ref{sec:astrophysics}. Section~\ref{sec:epoch_conclusions} summarizes the findings.

\section{Synergistic--Unique--Redundant Decomposition (SURD) for time series}
\label{sec:surd_method}
\subsection{Problem formulation}
Consider the observed stochastic processes
\begin{equation}
\mathbf{Q}(t)=\bigl(Q_1(t),Q_2(t),\ldots,Q_N(t)\bigr).
\end{equation}
For target index $j$, the future state is $Q_j^+\equiv Q_j(t+\Delta T)$. We write $Y\equiv Q_j^+$ and denote the observed predictor vector by $\mathbf{X}=(X_1,\ldots,X_M)$. Its entries may be different variables evaluated at different past times. In the present analysis $M=4$, with one past value of each variable, including the target's own history. Equivalently, after shifting the time origin, the target is $Y(t)=Q_j(t)$ and
\begin{equation}
X_i=Q_i(t-L-\delta_i),\qquad L=1,2,\ldots,30~{\rm d}.
\label{eq:predictors}
\end{equation}
The offsets $\delta_i$ define common or unequal lags. All predictors precede the target; no contemporaneous target is included.

\subsection{Information-theoretic preliminaries}
For discrete variables, entropy and conditional entropy are
\begin{align}
H(Y)&=-\sum_y p(y)\log_2p(y),\\
H(Y\mid\mathbf{X})&=-\sum_{y,\mathbf{x}}p(y,\mathbf{x})\log_2p(y\mid\mathbf{x}).
\end{align}
The total mutual information is
\begin{equation}
I(Y;\mathbf{X})=H(Y)-H(Y\mid\mathbf{X}),
\label{eq:mi_total}
\end{equation}
in bits \citep{Shannon1948,CoverThomas2006}. The unresolved part is the information leak,
\begin{equation}
\Delta I_{\mathrm{leak}\to j}=H(Y\mid\mathbf{X}),\qquad
H(Y)=I(Y;\mathbf{X})+\Delta I_{\mathrm{leak}\to j}.
\label{eq:forward_balance}
\end{equation}
This is an information balance. In observations, the unresolved term can include unobserved drivers, stochasticity, measurement error, inadequate histories and estimation effects; it is not exclusively a measure of hidden physical forcing.

\subsection{Unique, redundant and synergistic components}
SURD decomposes the mutual information as
\begin{equation}
I(Y;\mathbf{X})=
\sum_{\alpha=1}^{M}\Delta I^U_{\alpha\to j}
+\sum_{\mathcal{A}\in\mathcal{C}}\Delta I^R_{\mathcal{A}\to j}
+\sum_{\mathcal{A}\in\mathcal{C}}\Delta I^S_{\mathcal{A}\to j},
\label{eq:surd_master}
\end{equation}
where $\mathcal{C}$ contains all non-singleton predictor subsets. Unique information is available from one predictor and unavailable from any other individual predictor; redundant information is shared by the indicated predictors; synergistic information is obtained from their joint states. For four predictors there are four unique, eleven redundant and eleven synergistic terms, hence 26 components. A grouped redundant or synergistic term belongs to the group and cannot be counted as a separate full-strength directed link from every member.

The construction acts first on the specific information for each target state,
\begin{equation}
\tilde{\imath}_{\mathcal{A}}(y)=
\sum_{\mathbf{x}_{\mathcal{A}}}p(\mathbf{x}_{\mathcal{A}}\mid y)
\log_2\frac{p(\mathbf{x}_{\mathcal{A}}\mid y)}{p(\mathbf{x}_{\mathcal{A}})},
\end{equation}
and then averages the decomposed increments over $p(y)$ \citep{MartinezSanchez2024SURD,MartinezSanchez2025Thesis}. Thus SURD resolves different forms of multivariate predictability rather than only a single pairwise association. In particular, $\Delta I^U_{\alpha\to j}$ is not generally the conditional mutual information obtained by conditioning on all other predictors jointly. A vanishing unique term does not exclude the predictor from redundant or synergistic terms.

\subsection{Normalization and interpretation}
\label{sec:normalization}
Following the original notation, hats denote normalization of a component by the joint information. For brevity, $I_a$ denotes any one of the 26 terms in equation~\eqref{eq:surd_master}, and $f_a$ its hatted value. The component fraction, normalized leakage and target-entropy fraction are
\begin{equation}
\begin{aligned}
f_a&=\frac{I_a}{I(Y;\mathbf{X})}, &\sum_a f_a&=1,\\
\ell&=\widehat{\Delta I}_{\mathrm{leak}\to j}
=\frac{H(Y\mid\mathbf{X})}{H(Y)},\\
g_a&=\frac{I_a}{H(Y)}=(1-\ell)f_a,
&\sum_a g_a+\ell&=1.
\end{aligned}
\label{eq:normalizations}
\end{equation}
For example, $f_{U(F)}$ is $\widehat{\Delta I}^{U}_{F\to j}$ and $g_{U(F)}$ is the corresponding fraction of target entropy. We suppress the target and lag where the context is unambiguous. Fractions $f_a$ require $I(Y;\mathbf{X})>0$, as holds for the saved decompositions. When the true information is zero they are undefined, not evidence for a dominant interaction.

The first normalization identifies the composition of explained information; the third measures its magnitude relative to all target uncertainty. A component with $f_a=0.9$ and $\ell=0.99$ explains only $g_a=0.009$ of target entropy. Consequently, a large $f_a$ alone is insufficient to describe a strong influence. We report $f_a$, $\ell$ and $g_a$ at the same lag, and distinguish the lag maximizing $f_a$ from that maximizing $g_a$. Maxima are descriptive; neither a large fraction nor low leakage establishes statistical significance.

The primary ranking is performed separately for each named component: $L_{g,a}=\arg\max_{1\leq L\leq30}g_a(L)$. We then order distinct components within the unique, redundant or synergistic family by $g_a(L_{g,a})$, retaining the top three and their simultaneous $f_a$, $\ell$ and lag. Cross-variable unique terms are tabulated separately from target memory. This prevents a large redundant or self-history term from obscuring a distinguished synergy. Each listed component has its own maximizing lag; these maxima are not a decomposition at one common time and must not be summed. The accompanying catalogue also gives the $f_a$ maxima and rankings, and the complete components at selected common lags. An identically zero curve has no defined peak lag. Ranking is descriptive and imposes no significance threshold.

For the supplementary budget comparison in Appendix~\ref{app:budgets}, we define
\begin{equation}
\begin{aligned}
G_{U,\mathrm{self}}&=g_{U(j)},&
G_{U,\mathrm{cross}}&=\sum_{\alpha\ne j}g_{U(\alpha)},\\
G_R&=\sum_{\mathcal{A}}g_{R(\mathcal{A})},&
G_S&=\sum_{\mathcal{A}}g_{S(\mathcal{A})}.
\end{aligned}
\label{eq:budgets}
\end{equation}
These four budgets and $\ell$ sum to unity. An individual synergy need not be the largest component for the summed synergistic budget to be important. For information available from a single predictor, the exact SURD construction gives the identity \citep[equation~3.22]{MartinezSanchez2025Thesis}
\begin{equation}
\frac{I(Y;X_\alpha)}{H(Y)}
=g_{U(\alpha)}+\sum_{\mathcal{A}\ni\alpha}g_{R(\mathcal{A})}.
\label{eq:single_source}
\end{equation}
Equation~\eqref{eq:single_source} concerns observing $X_\alpha$ alone. Redundant information is accessible from that predictor even though other predictors also provide it. Synergistic information requires joint observations and therefore enters equation~\eqref{eq:surd_master}, not this single-source identity. A sum of all unique, redundant and synergistic terms involving $X_\alpha$ describes its participation in the full decomposition; it is neither $I(Y;X_\alpha)/H(Y)$ nor necessarily the information lost upon removing that predictor. These participation sums overlap and should not be added across sources.

\subsection{Practical estimation and causal scope}
Probabilities are estimated with joint histograms, $\hat p(y,\mathbf{x})=n(y,\mathbf{x})/N_{\rm tuples}$. Such discretized information estimators can be affected by finite-sample bias, particularly when the number of populated joint states becomes large relative to the available observations \citep{Paninski2003}. We use the unchanged archived implementation of SURD. Histogram state definitions, lag configurations and sampling are specified below. The new calculations operate on its saved components and leakage: no probability estimator, numerical tolerance or observational data reduction has been changed. The numerical and sampling diagnostics used to interpret these allocations are collected in Section~\ref{sec:limits}.

We use SURD as an observational causal-discovery framework: the named components identify candidate individual and joint influences, shared information, and unresolved driving. A modest component can be scientifically informative when its lag dependence and relative importance recur across state counts or predictor configurations. The causal interpretation assumes appropriate temporal ordering, a causal Markov description, faithfulness and sufficiently stationary statistics \citep{MartinezSanchez2025Thesis}. We treat each season separately and interpret the inferred influences within the specified variables and histories. The optical continuum is a proxy for the unmeasured
ionizing driver, rather than a direct measurement of it.
Multiwavelength reverberation-mapping observations of
NGC 5548 have demonstrated complex relationships
between continuum and emission-line variability
\citep{DeRosa2015}. Causal leakage is the normalized conditional entropy $\ell$; it measures unresolved information, rather than a luminosity or energy fraction.

Redundancy also guides model reduction \citep[section~4.10]{MartinezSanchez2025Thesis}. A large $R(C,R)$ means that either the core or red-wing history supplies that shared contribution to the chosen target. Their other unique and synergistic contributions determine whether both histories remain useful. For an entire predictor to be omitted, the relevant information loss is
\begin{equation}
\begin{aligned}
D_i&=I(Y;X_i\mid\mathbf X_{-i})\\
&=H(Y\mid\mathbf X_{-i})-H(Y\mid\mathbf X),\\
D_i/H(Y)&=\ell_{-i}-\ell,
\end{aligned}
\label{eq:predictor_omission}
\end{equation}
where $\mathbf X_{-i}$ retains the other predictors, on the same target-date sample. A zero loss permits omission without loss of predictive information; a small loss motivates a simpler model. The saved full-set components identify candidates for this test, while evaluating $D_i$ requires the underlying joint histogram or reduced-set information estimates.

\section{SURD components in the five historical NGC~5548 seasons}
\label{sec:epoch_surd_results}
\subsection{Series, epochs and analysis design}
The five separately analysed epochs are the historical campaigns labelled Years~1--5 in the compilation of \citet{Lu2022}; their boundaries follow \citet{Peterson2004ApJ}. We retain the original variable names: $H$ is total H$\beta$, $B$ its blue wing, $C$ its core, $R$ its red wing, and $F$ the 5100~\AA\ continuum. The \emph{total set} is $(H,B,C,R)$ and the \emph{F5100 set} is $(F,B,C,R)$. Each member is a target in turn, with all four members as lagged predictors.

The corrected line columns of \citet{WandersPeterson1996} define $B$, $C$ and $R$ over observed wavelengths 4870--4920, 4920--4960 and 4960--5010~\AA, respectively. These are fixed, unequal-flux spectral bands. The total and component fluxes come from the same spectra, with $H\simeq B+C+R$ up to the rounding of the published columns, so they are not independent measurements. Same-date observations are averaged. The continuum retains its host contribution; a constant additive host term does not change the within-season quantile labels used here \citep{Bentz2009}.

\begin{table}
\centering
\caption{Historical observing intervals and native support. $N_F$ and $N_{\rm line}$ count native continuum and spectroscopic dates. Grid tuples give the range over common-lag targets and both four-variable sets; the full two-set, three-configuration analysis retains 118--261 tuples. Interpolated tuples are not independent observations.}
\label{tab:support}
\scriptsize
\begin{tabular}{rlrrr}
\toprule
Epoch & JD$-2400000$ & $N_F$ & $N_{\rm line}$ & Grid tuples \\
\midrule
1989 & 47509--47809 & 124 & 48 & 134--196 \\
1990 & 47861--48179 & 94 & 38 & 161--232 \\
1991 & 48225--48534 & 65 & 32 & 133--211 \\
1992 & 48623--48898 & 83 & 46 & 211--244 \\
1993 & 48954--49255 & 191 & 60 & 231--261 \\
\bottomrule
\end{tabular}
\end{table}

Each series is interpolated onto a daily grid within its epoch. Linear interpolation does not extrapolate or bridge a bracketing gap longer than 15~d. The base lag $L$ in equation~\eqref{eq:predictors} spans 1--30 observed days in steps of one day. For common lags, every $\delta_i=0$. In the blue-leading configuration $\delta_B=5$~d; in the red-leading configuration $\delta_R=5$~d; other offsets remain zero. These names identify an imposed predictor arrangement, not a measured ordering of physical responses. If the shifted wing is itself the target, its self-history also moves from $L$ to $L+5$~d.

Two bins classify each flux as low or high relative to that variable's seasonal median. Three bins classify it as low, medium or high at its one-third and two-thirds marginal quantiles. Boundaries are obtained from the full native epoch series and shared by target and predictor roles. These are \emph{flux-amplitude states}, not velocity subdivisions, spatial BLR zones or time bins. State thresholds differ among variables and seasons, so a high state is relative to that season rather than a fixed luminosity. The joint histogram has $2^5=32$ or $3^5=243$ possible cells for four predictors and one target.

The complete design comprises $5\times2\times4\times3\times2\times30=7200$ decompositions, 187200 component values and 240 lag scans. The three public atlases contain 120 pages, each comparing the two bin choices for one epoch, set, target and configuration. All component sums agree with unity to $1.6\times10^{-15}$ in the saved results. The individual-component tables and new component figure are recalculated from these values and their leakage; all four original component plots and the earlier budget figure are retained; published comparison values are explicitly identified as such. No amplitude cut is used to select nonzero component peaks. A dash denotes a term that is zero throughout the scan, not a significance threshold.

\subsection{Common lags: named unique, synergistic and redundant influences}
\label{sec:common}
Tables~\ref{tab:cross_u_total_wings} and \ref{tab:cross_u_continuum_wings} list the three leading cross-variable unique components, Tables~\ref{tab:synergistic_total_wings} and \ref{tab:synergistic_continuum_wings} the leading individual synergies, and Tables~\ref{tab:redundant_total_wings} and \ref{tab:redundant_continuum_wings} the leading redundant groups. Every table covers all four targets and all five epochs, with two and three flux states shown separately. Each entry identifies the predictor or group, its $f_a$, $g_a$, leakage and lag. We rank within each family so that a physically interesting joint influence remains visible even when self-memory or a shared component is larger. The full-precision catalogue also records the alternative $f_a$ ranking.

Figures~\ref{fig:ep_common_1989_continuum_wings_core}--\ref{fig:ep_common_1993_total_wings_core} show the common-lag examples and display the full $f_a(L)$ curves and leakage. They reveal how several named components coexist and exchange relative importance as the lag changes. The unique self term reaches its largest $g_a$ at 1~d in all 80 common-lag target--bin scans; these memory terms are recorded separately in the accompanying catalogue. Appendix~\ref{app:dominant} retains the largest individual component of every common-lag scan, including self-memory.

\begin{SURDfigure}
\centering
\includegraphics[width=0.98\textwidth]{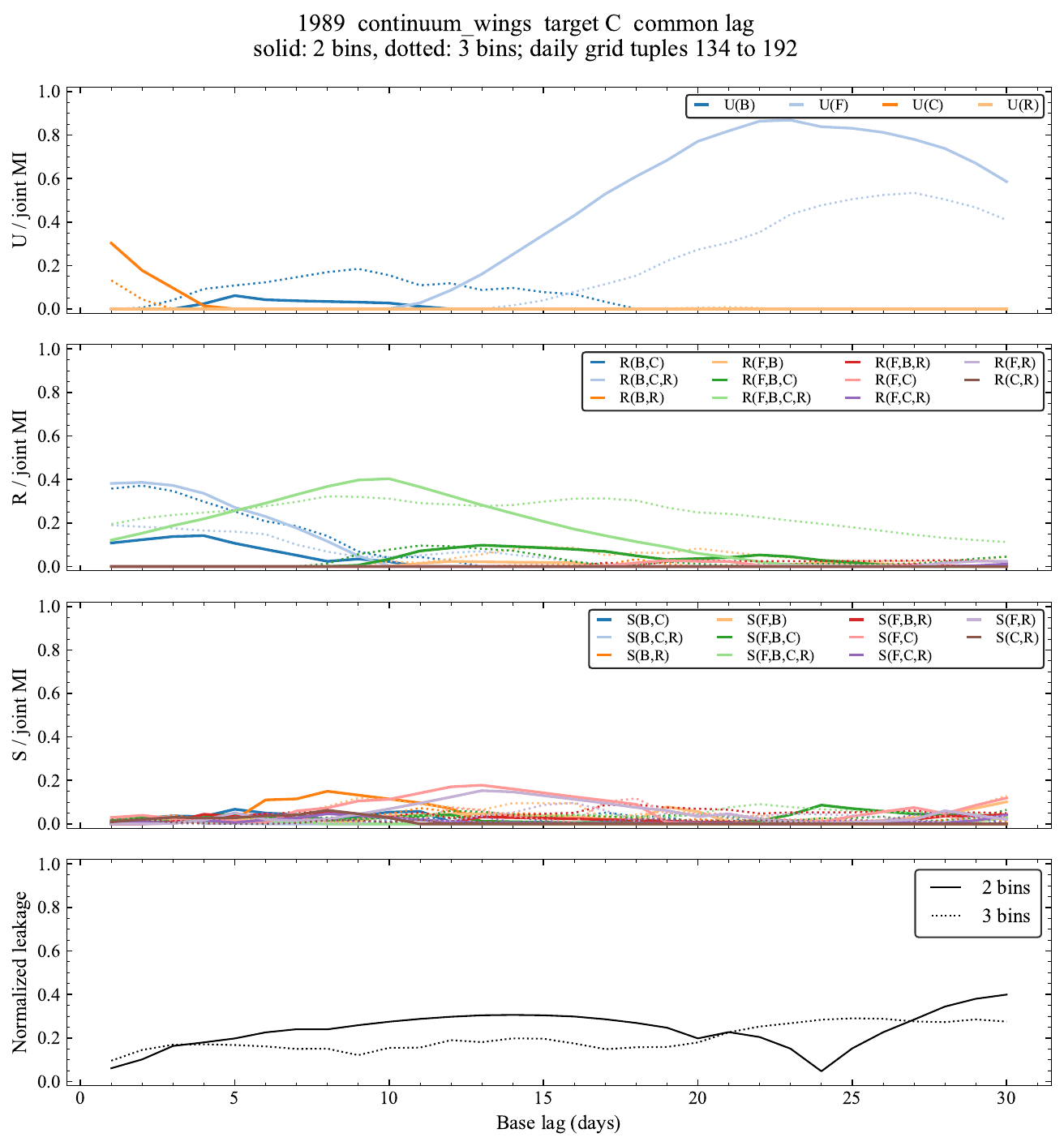}
\caption{Campaign 1989, F5100 set $(F,B,C,R)$, core target $C$, with common predictor lags. The 26 components are divided by the joint mutual information, $f_a=I_a/I(Y;\mathbf{X})$, in the U, R and S panels; the bottom panel shows normalized leakage $\ell=H(Y\mid\mathbf{X})/H(Y)$. Solid curves use two flux bins and dotted curves three bins. The daily grid supplies 134--192 tuples per lag. $F,B,C,R$ denote the continuum, blue wing, core and red wing. Source: common-lag atlas, page 3.}
\label{fig:ep_common_1989_continuum_wings_core}
\end{SURDfigure}
\begin{SURDfigure}
\centering
\includegraphics[width=0.98\textwidth]{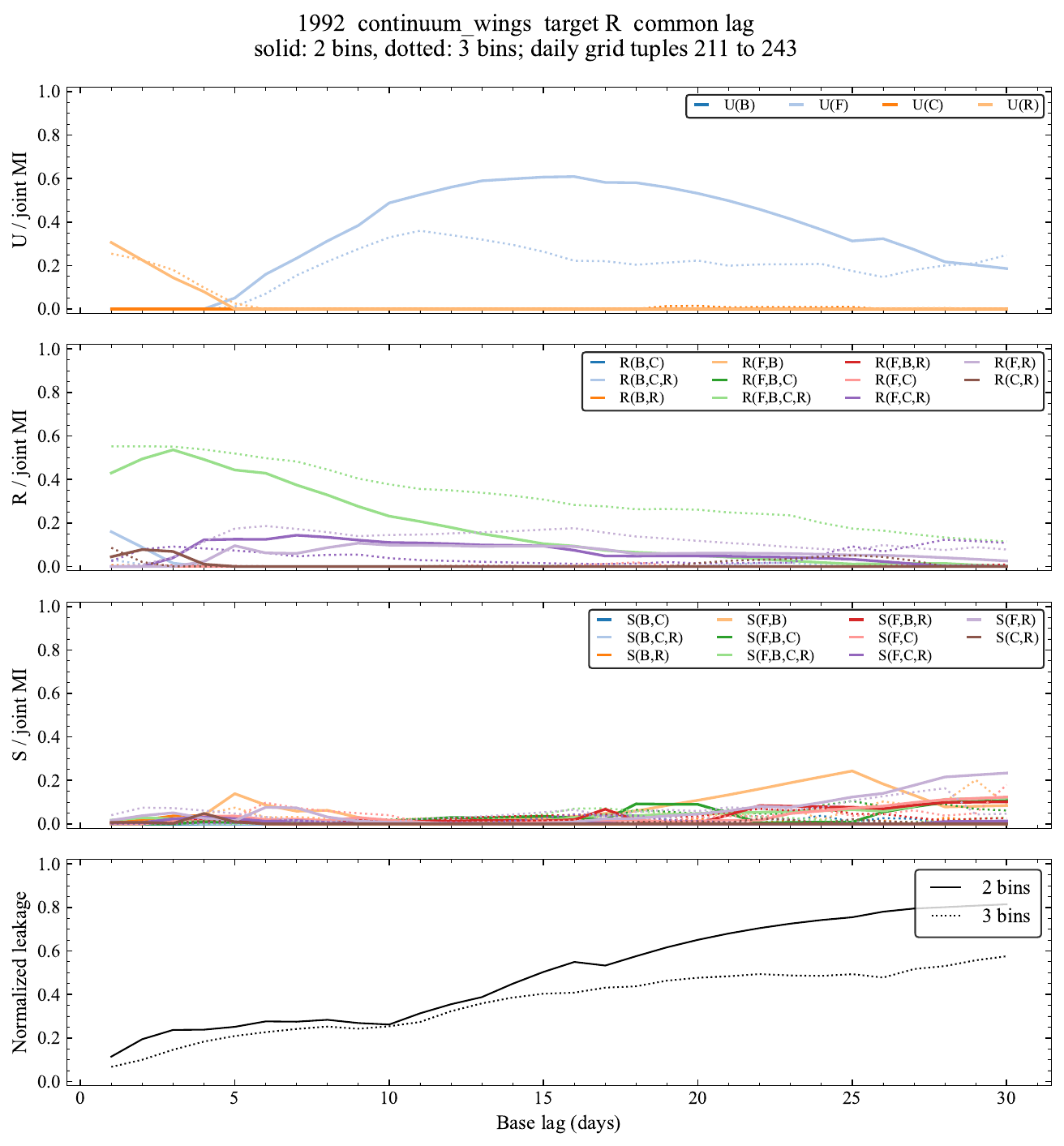}
\caption{Campaign 1992, F5100 set $(F,B,C,R)$, red-wing target $R$, with common predictor lags. The U, R and S panels show all 26 fractions $f_a=I_a/I(Y;\mathbf{X})$; the bottom panel shows $\ell=H(Y\mid\mathbf{X})/H(Y)$. Solid curves use two flux bins and dotted curves three bins; the daily grid supplies 211--243 tuples per lag. Symbols are as in Fig.~\ref{fig:ep_common_1989_continuum_wings_core}. Source: common-lag atlas, page 28.}
\label{fig:ep_common_1992_continuum_wings_red}
\end{SURDfigure}
\begin{SURDfigure}
\centering
\includegraphics[width=0.98\textwidth]{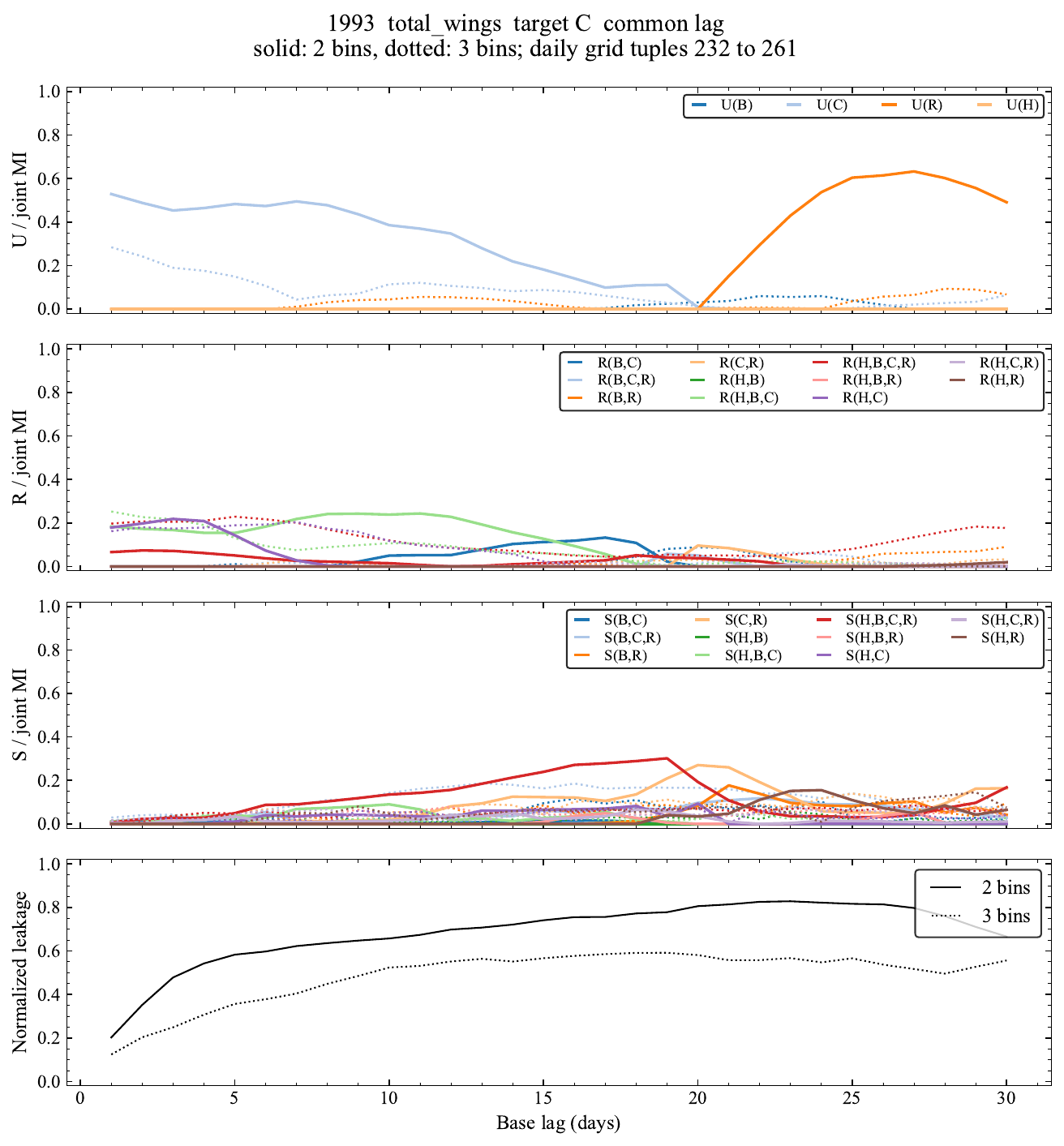}
\caption{Campaign 1993, total set $(H,B,C,R)$, core target $C$, with common predictor lags. The U, R and S panels show all 26 fractions $f_a=I_a/I(Y;\mathbf{X})$; the bottom panel shows $\ell=H(Y\mid\mathbf{X})/H(Y)$. Solid curves use two flux bins and dotted curves three bins; the daily grid supplies 232--261 tuples per lag. $H$ denotes total H$\beta$, and $B,C,R$ its blue wing, core and red wing. Source: common-lag atlas, page 38.}
\label{fig:ep_common_1993_total_wings_core}
\end{SURDfigure}

\paragraph*{Individual unique influences.}
In the F5100 set the continuum supplies the largest cross-variable unique component, ranked by maximum $g_a$, in 12 of the 15 epoch--line-target pairs under both bin choices. The exceptions are the 1991 blue and core targets and the 1993 red target. This is a recurring continuum-associated influence across the historical campaign, with substantial changes in its allocation between epochs and velocity bands.

The clearest case is $U(F)$ for the 1989 core (Fig.~\ref{fig:ep_common_1989_continuum_wings_core}). With two states, its maximum target-entropy contribution is $g=0.799$ at 24~d, with $f=0.838$ and $\ell=0.047$. With three states, $g=0.387$, $f=0.534$ and $\ell=0.276$ at 27~d. The maximum relative fraction in the two-state scan occurs one day earlier: $f=0.868$ at 23~d, with $g=0.738$ and $\ell=0.151$. Thus a distinguished continuum--core influence persists under both discretizations, and its estimated target-entropy contribution is quantified by pairing it with leakage at the same lag. Earlier in this scan, $R(B,C,R)$ and $U(C)$ contribute $g=0.359$ and 0.284 at 1~d with two states, while $S(F,C)$ reaches $g=0.124$ at 13~d with $\ell=0.304$. The delayed unique channel develops within this structured mixture of shared, self and joint information.

The 1992 red target provides another appreciable continuum influence (Fig.~\ref{fig:ep_common_1992_continuum_wings_red}). Its maximum $g_{U(F)}$ is 0.361 at 12~d with two states, with $\ell=0.355$, and 0.261 at 11~d with three, with $\ell=0.273$. The two-state maximum of $f_{U(F)}$ instead occurs at 16~d, where $f=0.609$, $g=0.274$ and $\ell=0.549$. The same figure shows a later $S(F,B)$ relative maximum of 0.244 at 25~d; its $g=0.060$ and $\ell=0.755$ identify a joint contribution within a largely unresolved late-time response.

The total set resolves unique information among the line components. In 1990, for example, $U(C)$ contributes $g=0.097$ to the red target at 5~d with two states and leakage 0.253. The red-wing history becomes especially informative in 1992: $U(R)\to B$ is nonzero in both sets and both bin choices, whereas the reverse $U(B)\to R$ curve is zero throughout all four common-lag scans. Table~\ref{tab:directed_1992} gives the corresponding extrema. In the F5100 set the red-to-blue maxima are $g=0.121/0.166$ at $5/9$~d with leakage $0.140/0.230$ for two/three states; in the total set they are $0.106/0.149$ at $10/9$~d with leakage $0.368/0.294$. The same red history supplies $g=0.102/0.181$ to future total H$\beta$ at $2/5$~d, with leakage $0.089/0.243$. These named components identify a consistent candidate red-to-blue influence and an informative red contribution to the subsequent integrated profile.

In 1993 the total-set core retains a distinct $U(R)$ contribution (Fig.~\ref{fig:ep_common_1993_total_wings_core}). With two states, $f=0.632$ at 27~d, $g=0.128$ and $\ell=0.797$; its largest $g=0.164$ occurs at 30~d with leakage 0.666. This is a resolved red-associated contribution in a target with a substantial unresolved budget. The core's own history is stronger at short lag, reaching $g=0.422/0.249$ at 1~d with leakage $0.202/0.124$. The separation of these named terms is informative even when no cross-variable component explains most of the target entropy.

\begin{table*}
\centering
\scriptsize
\caption{Ranked unique cross-series components for the total set $(H,B,C,R)$ with common lags. Each cell lists up to three distinct nonzero atoms in descending order of their own maximum $g_a=I_a/H(Y)$ over 1--30~d. Entries are $a:f_a/g_a/\ell\;(L)$, where $f_a=I_a/I(Y;\mathbf X)$ and leakage is evaluated at the same base lag $L$ in days. Different entries can have different lags and must not be added. $F,H,B,C,R$ denote continuum, total, blue, core and red fluxes; parentheses identify predictors, and the target is listed separately. Self-history is excluded. A dash means all eligible atoms are zero throughout the scan; smaller positive values that round to zero are printed as $<0.001$. Rankings are descriptive, without a significance threshold or uncertainty estimate.}
\label{tab:cross_u_total_wings}
\setlength{\tabcolsep}{5pt}
\renewcommand{\arraystretch}{1.15}
\begin{tabular}{ccp{0.39\textwidth}p{0.39\textwidth}}
\toprule
Epoch & Target & Two bins: $a:f_a/g_a/\ell\;(L)$ & Three bins: $a:f_a/g_a/\ell\;(L)$ \\
\midrule
1989 & $H$ & \shortstack[l]{$U(C):0.092/0.018/0.801\;(19)$ \\ $U(B):0.026/0.011/0.567\;(12)$ \\ $U(R):0.011/0.002/0.804\;(28)$} & \shortstack[l]{$U(B):0.134/0.089/0.332\;(9)$ \\ $U(C):0.029/0.011/0.630\;(28)$} \\
 & $B$ & \shortstack[l]{$U(H):0.060/0.046/0.230\;(7)$ \\ $U(C):0.015/0.009/0.410\;(11)$} & \shortstack[l]{$U(H):0.073/0.034/0.540\;(17)$ \\ $U(C):0.024/0.009/0.641\;(26)$} \\
 & $C$ & \shortstack[l]{$U(B):0.070/0.049/0.291\;(5)$ \\ $U(H):0.024/0.014/0.404\;(8)$ \\ $U(R):0.074/0.013/0.829\;(29)$} & \shortstack[l]{$U(B):0.154/0.108/0.299\;(9)$ \\ $U(H):0.050/0.041/0.180\;(3)$} \\
 & $R$ & \shortstack[l]{$U(H):0.126/0.071/0.440\;(6)$ \\ $U(B):0.125/0.028/0.776\;(16)$ \\ $U(C):0.038/0.011/0.697\;(12)$} & \shortstack[l]{$U(B):0.142/0.069/0.517\;(9)$ \\ $U(H):0.135/0.068/0.497\;(21)$ \\ $U(C):0.034/0.017/0.509\;(25)$} \\
\addlinespace[3pt]
1990 & $H$ & \shortstack[l]{$U(C):0.111/0.096/0.133\;(5)$ \\ $U(R):0.020/0.020/0.039\;(3)$ \\ $U(B):0.104/0.010/0.908\;(29)$} & \shortstack[l]{$U(R):0.130/0.055/0.574\;(20)$ \\ $U(B):0.033/0.028/0.172\;(4)$ \\ $U(C):0.008/0.006/0.272\;(6)$} \\
 & $B$ & \shortstack[l]{$U(H):0.097/0.079/0.187\;(5)$ \\ $U(C):0.134/0.078/0.413\;(10)$ \\ $U(R):0.006/0.004/0.330\;(8)$} & \shortstack[l]{$U(R):0.097/0.043/0.553\;(20)$ \\ $U(H):0.048/0.020/0.584\;(27)$ \\ $U(C):0.024/0.009/0.611\;(30)$} \\
 & $C$ & \shortstack[l]{$U(B):0.143/0.014/0.905\;(29)$} & \shortstack[l]{$U(R):0.098/0.043/0.560\;(21)$ \\ $U(B):0.037/0.029/0.213\;(6)$ \\ $U(H):0.004/0.004/0.086\;(2)$} \\
 & $R$ & \shortstack[l]{$U(C):0.130/0.097/0.253\;(5)$ \\ $U(B):0.065/0.005/0.922\;(30)$} & \shortstack[l]{$U(B):0.141/0.047/0.668\;(26)$ \\ $U(C):0.015/0.010/0.290\;(4)$ \\ $U(H):0.013/0.007/0.436\;(7)$} \\
\addlinespace[3pt]
1991 & $H$ & \shortstack[l]{$U(C):0.122/0.075/0.389\;(10)$ \\ $U(R):0.055/0.047/0.139\;(5)$} & \shortstack[l]{$U(B):0.112/0.097/0.134\;(5)$ \\ $U(R):0.115/0.070/0.387\;(29)$ \\ $U(C):0.021/0.013/0.401\;(23)$} \\
 & $B$ & \shortstack[l]{$U(C):0.135/0.081/0.395\;(12)$ \\ $U(H):0.033/0.031/0.073\;(2)$ \\ $U(R):0.034/0.030/0.137\;(7)$} & \shortstack[l]{$U(R):0.124/0.084/0.322\;(30)$ \\ $U(H):0.035/0.030/0.132\;(5)$ \\ $U(C):0.018/0.012/0.329\;(24)$} \\
 & $C$ & $\text{---}$ & \shortstack[l]{$U(B):0.134/0.118/0.121\;(6)$ \\ $U(R):0.120/0.076/0.365\;(29)$ \\ $U(H):0.032/0.030/0.064\;(3)$} \\
 & $R$ & \shortstack[l]{$U(C):0.158/0.055/0.653\;(10)$} & \shortstack[l]{$U(B):0.295/0.129/0.562\;(11)$ \\ $U(H):0.036/0.016/0.561\;(18)$ \\ $U(C):0.020/0.009/0.561\;(27)$} \\
\addlinespace[3pt]
1992 & $H$ & \shortstack[l]{$U(R):0.111/0.102/0.089\;(2)$} & \shortstack[l]{$U(R):0.239/0.181/0.243\;(5)$ \\ $U(C):0.018/0.006/0.670\;(27)$} \\
 & $B$ & \shortstack[l]{$U(R):0.168/0.106/0.368\;(10)$ \\ $U(H):0.049/0.044/0.100\;(2)$} & \shortstack[l]{$U(R):0.212/0.149/0.294\;(9)$ \\ $U(C):0.016/0.005/0.674\;(28)$ \\ $U(H):0.001/0.001/0.076\;(3)$} \\
 & $C$ & \shortstack[l]{$U(R):0.201/0.073/0.636\;(12)$} & \shortstack[l]{$U(R):0.234/0.159/0.321\;(7)$ \\ $U(H):0.024/0.021/0.114\;(2)$} \\
 & $R$ & $\text{---}$ & \shortstack[l]{$U(C):0.015/0.005/0.668\;(25)$} \\
\addlinespace[3pt]
1993 & $H$ & \shortstack[l]{$U(R):0.295/0.078/0.735\;(30)$ \\ $U(C):0.215/0.071/0.669\;(11)$} & \shortstack[l]{$U(C):0.154/0.098/0.364\;(5)$ \\ $U(R):0.100/0.042/0.581\;(28)$ \\ $U(B):0.013/0.006/0.564\;(26)$} \\
 & $B$ & \shortstack[l]{$U(H):0.318/0.133/0.582\;(5)$ \\ $U(C):0.241/0.080/0.669\;(15)$ \\ $U(R):0.421/0.040/0.905\;(30)$} & \shortstack[l]{$U(C):0.161/0.086/0.463\;(10)$ \\ $U(R):0.059/0.030/0.485\;(16)$ \\ $U(H):0.030/0.020/0.344\;(6)$} \\
 & $C$ & \shortstack[l]{$U(R):0.491/0.164/0.666\;(30)$} & \shortstack[l]{$U(R):0.093/0.047/0.496\;(28)$ \\ $U(B):0.059/0.027/0.548\;(24)$} \\
 & $R$ & \shortstack[l]{$U(H):0.329/0.083/0.749\;(30)$ \\ $U(B):0.136/0.031/0.776\;(17)$} & \shortstack[l]{$U(H):0.113/0.047/0.581\;(29)$ \\ $U(B):0.108/0.042/0.614\;(19)$ \\ $U(C):0.031/0.013/0.592\;(24)$} \\
\bottomrule
\end{tabular}
\end{table*}

\begin{table*}
\centering
\scriptsize
\caption{Ranked unique cross-series components for the F5100 set $(F,B,C,R)$ with common lags. Notation and independent peak selection follow Table~\ref{tab:cross_u_total_wings}. Self-history is excluded. The three entries in a cell need not share a lag and must not be added. No significance threshold is applied.}
\label{tab:cross_u_continuum_wings}
\setlength{\tabcolsep}{5pt}
\renewcommand{\arraystretch}{1.15}
\begin{tabular}{ccp{0.39\textwidth}p{0.39\textwidth}}
\toprule
Epoch & Target & Two bins: $a:f_a/g_a/\ell\;(L)$ & Three bins: $a:f_a/g_a/\ell\;(L)$ \\
\midrule
1989 & $F$ & \shortstack[l]{$U(B):0.370/0.129/0.651\;(30)$} & \shortstack[l]{$U(B):0.076/0.040/0.467\;(27)$ \\ $U(C):0.023/0.013/0.424\;(30)$ \\ $U(R):0.022/0.013/0.415\;(16)$} \\
 & $B$ & \shortstack[l]{$U(F):0.721/0.455/0.370\;(26)$ \\ $U(C):0.010/0.009/0.122\;(8)$} & \shortstack[l]{$U(F):0.475/0.365/0.233\;(27)$} \\
 & $C$ & \shortstack[l]{$U(F):0.838/0.799/0.047\;(24)$ \\ $U(B):0.062/0.049/0.198\;(5)$} & \shortstack[l]{$U(F):0.534/0.387/0.276\;(27)$ \\ $U(B):0.185/0.162/0.121\;(9)$ \\ $U(R):0.008/0.006/0.226\;(21)$} \\
 & $R$ & \shortstack[l]{$U(F):0.796/0.598/0.248\;(22)$ \\ $U(C):0.072/0.039/0.468\;(10)$} & \shortstack[l]{$U(F):0.313/0.203/0.351\;(19)$ \\ $U(B):0.098/0.063/0.359\;(9)$ \\ $U(C):0.013/0.008/0.365\;(28)$} \\
\addlinespace[3pt]
1990 & $F$ & \shortstack[l]{$U(R):0.130/0.048/0.633\;(30)$ \\ $U(C):0.011/0.004/0.650\;(24)$} & \shortstack[l]{$U(C):0.173/0.092/0.471\;(30)$ \\ $U(R):0.024/0.013/0.470\;(29)$ \\ $U(B):0.009/0.005/0.464\;(11)$} \\
 & $B$ & \shortstack[l]{$U(F):0.788/0.480/0.390\;(20)$ \\ $U(C):0.102/0.079/0.228\;(10)$ \\ $U(R):0.007/0.006/0.143\;(6)$} & \shortstack[l]{$U(F):0.352/0.270/0.232\;(19)$ \\ $U(R):0.073/0.052/0.294\;(29)$} \\
 & $C$ & \shortstack[l]{$U(F):0.579/0.311/0.463\;(19)$} & \shortstack[l]{$U(F):0.305/0.238/0.220\;(21)$ \\ $U(B):0.051/0.046/0.111\;(6)$ \\ $U(R):0.031/0.023/0.245\;(19)$} \\
 & $R$ & \shortstack[l]{$U(F):0.567/0.264/0.534\;(25)$ \\ $U(C):0.129/0.100/0.227\;(5)$} & \shortstack[l]{$U(F):0.353/0.216/0.388\;(19)$ \\ $U(B):0.020/0.016/0.195\;(5)$ \\ $U(C):0.012/0.010/0.174\;(4)$} \\
\addlinespace[3pt]
1991 & $F$ & \shortstack[l]{$U(B):0.342/0.086/0.748\;(24)$ \\ $U(C):0.141/0.031/0.777\;(15)$ \\ $U(R):0.009/0.003/0.707\;(30)$} & \shortstack[l]{$U(R):0.189/0.098/0.481\;(24)$ \\ $U(B):0.138/0.050/0.638\;(12)$ \\ $U(C):0.005/0.003/0.492\;(7)$} \\
 & $B$ & \shortstack[l]{$U(F):0.287/0.230/0.198\;(23)$ \\ $U(C):0.153/0.113/0.259\;(12)$ \\ $U(R):0.032/0.030/0.074\;(7)$} & \shortstack[l]{$U(R):0.107/0.076/0.297\;(29)$ \\ $U(C):0.031/0.021/0.317\;(23)$ \\ $U(F):0.009/0.006/0.319\;(21)$} \\
 & $C$ & $\text{---}$ & \shortstack[l]{$U(B):0.181/0.167/0.080\;(6)$ \\ $U(R):0.111/0.078/0.297\;(29)$ \\ $U(F):0.036/0.024/0.316\;(20)$} \\
 & $R$ & \shortstack[l]{$U(F):0.440/0.250/0.433\;(18)$ \\ $U(C):0.059/0.021/0.641\;(27)$} & \shortstack[l]{$U(F):0.227/0.154/0.321\;(13)$ \\ $U(B):0.102/0.056/0.449\;(26)$ \\ $U(C):0.016/0.009/0.470\;(27)$} \\
\addlinespace[3pt]
1992 & $F$ & $\text{---}$ & \shortstack[l]{$U(R):0.052/0.023/0.564\;(20)$ \\ $U(C):0.044/0.022/0.487\;(14)$} \\
 & $B$ & \shortstack[l]{$U(F):0.556/0.286/0.485\;(18)$ \\ $U(R):0.140/0.121/0.140\;(5)$} & \shortstack[l]{$U(F):0.263/0.170/0.354\;(19)$ \\ $U(R):0.216/0.166/0.230\;(9)$ \\ $U(C):0.011/0.005/0.532\;(28)$} \\
 & $C$ & \shortstack[l]{$U(F):0.384/0.209/0.454\;(14)$ \\ $U(R):0.035/0.027/0.227\;(5)$} & \shortstack[l]{$U(F):0.231/0.170/0.264\;(12)$ \\ $U(R):0.200/0.157/0.216\;(6)$} \\
 & $R$ & \shortstack[l]{$U(F):0.560/0.361/0.355\;(12)$} & \shortstack[l]{$U(F):0.360/0.261/0.273\;(11)$ \\ $U(C):0.014/0.007/0.477\;(20)$} \\
\addlinespace[3pt]
1993 & $F$ & \shortstack[l]{$U(R):0.273/0.071/0.739\;(21)$ \\ $U(C):0.023/0.005/0.799\;(27)$} & \shortstack[l]{$U(C):0.103/0.056/0.459\;(25)$ \\ $U(B):0.010/0.006/0.403\;(18)$ \\ $U(R):0.007/0.004/0.379\;(20)$} \\
 & $B$ & \shortstack[l]{$U(F):0.613/0.211/0.656\;(26)$ \\ $U(C):0.256/0.085/0.668\;(7)$} & \shortstack[l]{$U(F):0.321/0.171/0.467\;(27)$ \\ $U(C):0.154/0.108/0.299\;(10)$} \\
 & $C$ & \shortstack[l]{$U(F):0.442/0.198/0.551\;(21)$ \\ $U(R):0.083/0.038/0.536\;(30)$} & \shortstack[l]{$U(F):0.396/0.246/0.379\;(20)$ \\ $U(R):0.063/0.044/0.306\;(28)$ \\ $U(B):0.038/0.024/0.360\;(24)$} \\
 & $R$ & \shortstack[l]{$U(B):0.251/0.062/0.754\;(18)$ \\ $U(C):0.009/0.003/0.673\;(27)$} & \shortstack[l]{$U(F):0.147/0.068/0.540\;(12)$ \\ $U(B):0.123/0.061/0.502\;(19)$ \\ $U(C):0.034/0.017/0.502\;(19)$} \\
\bottomrule
\end{tabular}
\end{table*}

\begin{SURDtable}
\centering
\caption{Unique red-wing contributions in 1992 with common lags. Values refer to the maximum $g_a$ of the named component, with simultaneous $f_a$ and leakage. Lags are observed days. These resolve the red-to-blue channel in both subsets and the red-to-total channel in the total set.}
\label{tab:directed_1992}
\small
\begin{tabular}{lcccrrrr}
\toprule
Set & Target & Bins & Component & $L$ & $f$ & $g$ & $\ell$ \\
\midrule
F5100 & $B$ & 2 & $U(R)$ & 5 & 0.140 & 0.121 & 0.140 \\
F5100 & $B$ & 3 & $U(R)$ & 9 & 0.216 & 0.166 & 0.230 \\
Total & $B$ & 2 & $U(R)$ & 10 & 0.168 & 0.106 & 0.368 \\
Total & $B$ & 3 & $U(R)$ & 9 & 0.212 & 0.149 & 0.294 \\
Total & $H$ & 2 & $U(R)$ & 2 & 0.111 & 0.102 & 0.089 \\
Total & $H$ & 3 & $U(R)$ & 5 & 0.239 & 0.181 & 0.243 \\
\bottomrule
\end{tabular}
\end{SURDtable}

\paragraph*{Individual synergistic influences.}
The pair or group responsible for synergy changes with season. In the 1989 F5100-set blue target, $S(F,R)$ reaches $g=0.193$ at 16~d with two states and $\ell=0.267$; with three states it reaches 0.091 at 17~d with $\ell=0.223$. The agreement of the named pair and nearby peak lags identifies a candidate joint continuum--red-wing influence on the blue wing. In the total set, the same blue target instead has $S(B,R)$ among its leading synergies: its two-state peak is $g=0.163$ at 6~d with leakage 0.220. This term combines blue-wing memory with the earlier red-wing state.

In 1990, two distinct synergies involving the continuum are prominent for the core. The two-state maxima of $S(F,B)$ and $S(F,C)$ are respectively $g=0.150$ at 12~d and 0.128 at 13~d, with leakage 0.399 and 0.409. With three states the same pairs reach 0.085 at 13~d and 0.115 at 12~d, with leakage 0.221 and 0.220. Their persistence near 12--13~d provides a useful joint-response signature even though the components individually account for only a moderate fraction of target entropy. For the blue target, $S(F,B)$ is stronger with three states: $g=0.104$ at 13~d and 0.224 at 12~d for two/three states, with leakage 0.305 and 0.263. Here the continuum and the target's own history form a particularly informative joint predictor.

The 1991 core is a particularly clear demonstration of why named synergies must accompany unique terms. Its two-state $U(F)$ vanishes at all common lags, yet at 20~d $S(F,B)$ and $S(F,R)$ contribute $g=0.187$ and 0.125, while $R(F,B,C,R)$ contributes 0.327 and leakage is only 0.087. These three specific components locate the shared and joint continuum information directly. With three states, the separate maxima of $S(F,B)$ and $S(F,R)$ are 0.061 at 15~d and 0.085 at 30~d, with leakage 0.190 and 0.327. The identities persist while their preferred delays and amplitudes change. In the total-set blue target, $S(C,R)$ reaches almost the same $g=0.113$ with both state counts, but at 7 and 24~d, respectively, with leakage 0.137 and 0.329. This identifies the pair while retaining the considerable lag sensitivity.

For the 1992 F5100-set core, $S(F,B)$ reaches $g=0.121$ at 7~d with two states and leakage 0.239. Its three-state maximum is 0.052 at 17~d with leakage 0.382. In 1993 the distinguished core synergy instead involves the red wing: $S(F,R)$ reaches $g=0.172$ at 28~d with two states and $g=0.076$ at 29~d with three, with leakage 0.570 and 0.343. The similar late peak locations make this an informative candidate joint influence despite incomplete prediction of the core. In the total set, the 1993 blue target also displays a three-variable $S(B,C,R)$ contribution, with $g=0.063/0.105$ at $24/23$~d and leakage $0.843/0.545$. The full rankings in Tables~\ref{tab:synergistic_total_wings} and \ref{tab:synergistic_continuum_wings} retain these moderate components alongside their larger or smaller competitors.

\begin{figure*}
\centering
\includegraphics[width=0.98\textwidth]{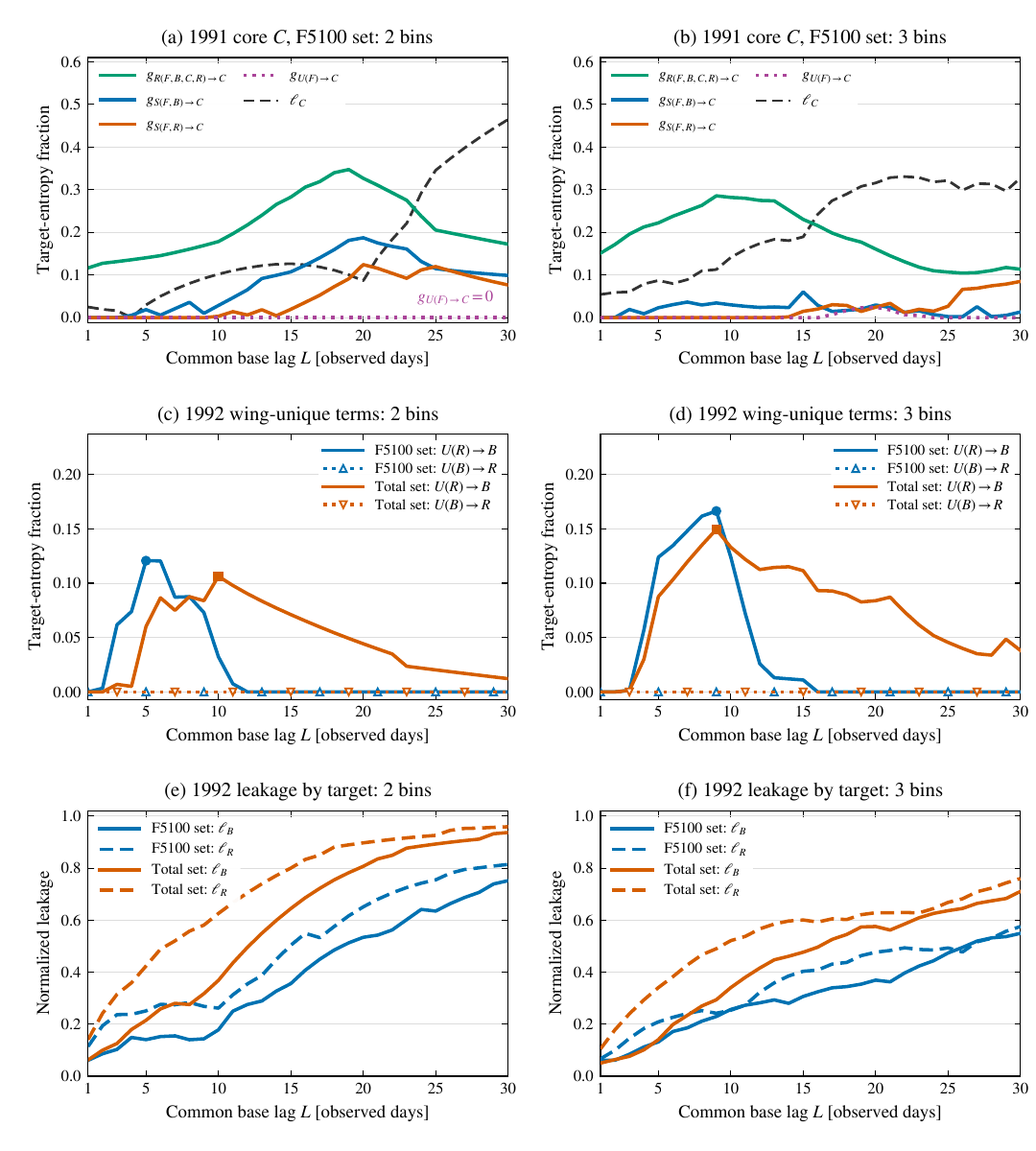}
\caption{Named SURD components and target-specific leakage for common predictor
lags. The columns compare two and three flux states.
Panels (a,b) show four individual contributions to the 1991 core target in the
F5100 set: $R(F,B,C,R)$, $S(F,B)$, $S(F,R)$ and $U(F)$, together with that target's
leakage $\ell_C$ (black dashed). The joint and shared terms remain nonzero when
the two-state unique continuum term vanishes.
Panels (c,d) compare the individual unique terms $U(R)\to B$ and $U(B)\to R$ in
both variable sets. Filled markers identify the maxima of the nonzero curves;
alternating open markers distinguish the two reverse-direction curves, which
are identically zero throughout these scans.
Panels (e,f) supply the corresponding leakage for the blue target (solid) and
red target (dashed); colours distinguish the F5100 and total sets.
Every component ordinate is its fraction of target entropy,
$g_a=(1-\ell)f_a$; leakage is evaluated for the same target, variable set, bin
choice and lag. Here $F$ is the optical continuum, $B$ the blue wing, $C$ the
core and $R$ the red wing. The individual components and their simultaneous leakage identify the resolved pathways discussed in the text.}
\label{fig:named_components}
\end{figure*}

Figure~\ref{fig:named_components} displays the individual 1991 core terms and the 1992 directional red--blue comparison. Its upper panels show continuum-containing synergy even where $U(F)$ vanishes. Its middle panels isolate the $U(R)\to B$ channel in both sets, and its lower panels give the leakage for each corresponding target and state count. Thus the directional comparison can be read together with the completeness of each target's information budget.

\begin{table*}
\centering
\scriptsize
\caption{Ranked synergistic components for the total set $(H,B,C,R)$ with common lags. Notation and independent peak selection follow Table~\ref{tab:cross_u_total_wings}. The three entries in a cell need not share a lag and must not be added. No significance threshold is applied.}
\label{tab:synergistic_total_wings}
\setlength{\tabcolsep}{5pt}
\renewcommand{\arraystretch}{1.15}
\begin{tabular}{ccp{0.39\textwidth}p{0.39\textwidth}}
\toprule
Epoch & Target & Two bins: $a:f_a/g_a/\ell\;(L)$ & Three bins: $a:f_a/g_a/\ell\;(L)$ \\
\midrule
1989 & $H$ & \shortstack[l]{$S(B,R):0.449/0.085/0.810\;(27)$ \\ $S(H,B):0.134/0.080/0.403\;(8)$ \\ $S(B,C):0.121/0.052/0.567\;(12)$} & \shortstack[l]{$S(H,R):0.099/0.040/0.595\;(23)$ \\ $S(H,B):0.083/0.039/0.533\;(19)$ \\ $S(B,C):0.075/0.039/0.486\;(17)$} \\
 & $B$ & \shortstack[l]{$S(B,R):0.209/0.163/0.220\;(6)$ \\ $S(H,B):0.191/0.147/0.230\;(7)$ \\ $S(B,C):0.264/0.071/0.733\;(21)$} & \shortstack[l]{$S(B,R):0.096/0.045/0.531\;(18)$ \\ $S(B,C):0.071/0.041/0.419\;(12)$ \\ $S(H,B):0.047/0.041/0.134\;(3)$} \\
 & $C$ & \shortstack[l]{$S(H,B):0.125/0.075/0.404\;(8)$ \\ $S(B,R):0.382/0.071/0.815\;(30)$ \\ $S(B,C,R):0.276/0.054/0.806\;(27)$} & \shortstack[l]{$S(B,R):0.104/0.056/0.462\;(19)$ \\ $S(B,C):0.068/0.043/0.375\;(13)$ \\ $S(H,B):0.081/0.040/0.507\;(21)$} \\
 & $R$ & \shortstack[l]{$S(B,R):0.157/0.039/0.751\;(15)$ \\ $S(B,C):0.132/0.036/0.727\;(14)$ \\ $S(B,C,R):0.285/0.034/0.879\;(23)$} & \shortstack[l]{$S(H,R):0.122/0.062/0.497\;(21)$ \\ $S(B,R):0.106/0.057/0.459\;(22)$ \\ $S(B,C):0.092/0.043/0.530\;(16)$} \\
\addlinespace[3pt]
1990 & $H$ & \shortstack[l]{$S(H,R):0.063/0.060/0.039\;(3)$ \\ $S(H,C,R):0.050/0.043/0.133\;(5)$ \\ $S(B,C):0.042/0.039/0.078\;(4)$} & \shortstack[l]{$S(B,C,R):0.170/0.069/0.594\;(22)$ \\ $S(B,R):0.067/0.040/0.405\;(9)$ \\ $S(B,C):0.100/0.037/0.630\;(24)$} \\
 & $B$ & \shortstack[l]{$S(H,B):0.051/0.044/0.154\;(3)$ \\ $S(B,C,R):0.315/0.041/0.871\;(25)$ \\ $S(H,B,C,R):0.050/0.040/0.187\;(5)$} & \shortstack[l]{$S(B,C,R):0.141/0.063/0.552\;(21)$ \\ $S(B,R):0.092/0.053/0.431\;(9)$ \\ $S(B,C):0.109/0.043/0.608\;(25)$} \\
 & $C$ & \shortstack[l]{$S(H,C,R):0.051/0.036/0.291\;(5)$ \\ $S(H,C):0.040/0.034/0.140\;(2)$ \\ $S(B,C,R):0.328/0.031/0.905\;(29)$} & \shortstack[l]{$S(B,C):0.122/0.050/0.591\;(26)$ \\ $S(B,R):0.081/0.048/0.410\;(11)$ \\ $S(B,C,R):0.101/0.045/0.560\;(21)$} \\
 & $R$ & \shortstack[l]{$S(C,R):0.067/0.059/0.114\;(2)$ \\ $S(H,C,R):0.046/0.034/0.253\;(5)$ \\ $S(B,C,R):0.164/0.026/0.841\;(24)$} & \shortstack[l]{$S(B,C,R):0.186/0.068/0.633\;(23)$ \\ $S(H,B):0.102/0.033/0.675\;(27)$ \\ $S(B,R):0.046/0.033/0.290\;(4)$} \\
\addlinespace[3pt]
1991 & $H$ & \shortstack[l]{$S(C,R):0.104/0.093/0.107\;(4)$ \\ $S(B,R):0.059/0.053/0.107\;(4)$ \\ $S(H,C,R):0.040/0.032/0.187\;(6)$} & \shortstack[l]{$S(B,R):0.178/0.111/0.376\;(30)$ \\ $S(H,R):0.129/0.081/0.371\;(27)$ \\ $S(B,C,R):0.109/0.068/0.379\;(26)$} \\
 & $B$ & \shortstack[l]{$S(C,R):0.131/0.113/0.137\;(7)$ \\ $S(B,R):0.079/0.071/0.106\;(5)$ \\ $S(H,C,R):0.045/0.038/0.137\;(7)$} & \shortstack[l]{$S(B,R):0.209/0.139/0.335\;(23)$ \\ $S(H,R):0.172/0.116/0.322\;(30)$ \\ $S(C,R):0.169/0.113/0.329\;(24)$} \\
 & $C$ & \shortstack[l]{$S(C,R):0.093/0.090/0.028\;(4)$ \\ $S(B,C):0.026/0.025/0.027\;(2)$ \\ $S(H,C):0.025/0.024/0.027\;(2)$} & \shortstack[l]{$S(B,R):0.165/0.106/0.359\;(26)$ \\ $S(B,C,R):0.128/0.081/0.370\;(24)$ \\ $S(H,R):0.109/0.069/0.365\;(29)$} \\
 & $R$ & \shortstack[l]{$S(H,C,R):0.142/0.077/0.459\;(6)$ \\ $S(C,R):0.100/0.047/0.533\;(7)$ \\ $S(H,C):0.090/0.031/0.653\;(10)$} & \shortstack[l]{$S(B,R):0.230/0.106/0.538\;(24)$ \\ $S(H,R):0.175/0.085/0.516\;(26)$ \\ $S(C,R):0.142/0.065/0.538\;(24)$} \\
\addlinespace[3pt]
1992 & $H$ & \shortstack[l]{$S(B,R):0.078/0.075/0.042\;(1)$ \\ $S(C,R):0.080/0.060/0.248\;(5)$ \\ $S(B,C,R):0.028/0.023/0.176\;(3)$} & \shortstack[l]{$S(H,R):0.099/0.042/0.577\;(20)$ \\ $S(B,R):0.100/0.038/0.625\;(22)$ \\ $S(H,B,R):0.087/0.037/0.577\;(20)$} \\
 & $B$ & \shortstack[l]{$S(C,R):0.089/0.065/0.275\;(8)$ \\ $S(B,C):0.046/0.044/0.062\;(1)$ \\ $S(B,R):0.041/0.035/0.126\;(3)$} & \shortstack[l]{$S(H,B):0.053/0.050/0.064\;(2)$ \\ $S(B,R):0.049/0.045/0.076\;(3)$ \\ $S(H,R):0.099/0.041/0.585\;(22)$} \\
 & $C$ & \shortstack[l]{$S(C,R):0.064/0.044/0.316\;(4)$ \\ $S(B,R):0.034/0.029/0.159\;(2)$ \\ $S(H,C):0.024/0.022/0.083\;(1)$} & \shortstack[l]{$S(B,R):0.098/0.039/0.597\;(21)$ \\ $S(H,B,R):0.084/0.036/0.574\;(19)$ \\ $S(H,R):0.089/0.032/0.641\;(25)$} \\
 & $R$ & \shortstack[l]{$S(C,R):0.057/0.037/0.359\;(4)$ \\ $S(H,R):0.027/0.018/0.315\;(3)$ \\ $S(B,C,R):0.041/0.017/0.581\;(9)$} & \shortstack[l]{$S(B,R):0.131/0.049/0.628\;(20)$ \\ $S(H,R):0.103/0.042/0.593\;(16)$ \\ $S(H,B,R):0.088/0.035/0.602\;(18)$} \\
\addlinespace[3pt]
1993 & $H$ & \shortstack[l]{$S(H,B,C,R):0.235/0.061/0.741\;(16)$ \\ $S(B,C,R):0.183/0.061/0.669\;(11)$ \\ $S(H,C):0.114/0.051/0.550\;(4)$} & \shortstack[l]{$S(B,C,R):0.215/0.097/0.546\;(21)$ \\ $S(B,R):0.141/0.065/0.541\;(23)$ \\ $S(H,R):0.136/0.063/0.538\;(15)$} \\
 & $B$ & \shortstack[l]{$S(H,B,C,R):0.275/0.091/0.668\;(19)$ \\ $S(B,C,R):0.403/0.063/0.843\;(24)$ \\ $S(C,R):0.298/0.055/0.815\;(23)$} & \shortstack[l]{$S(B,C,R):0.231/0.105/0.545\;(23)$ \\ $S(C,R):0.145/0.076/0.477\;(13)$ \\ $S(H,B,C,R):0.155/0.068/0.561\;(29)$} \\
 & $C$ & \shortstack[l]{$S(H,B,C,R):0.279/0.068/0.756\;(17)$ \\ $S(C,R):0.164/0.055/0.666\;(30)$ \\ $S(B,R):0.177/0.033/0.814\;(21)$} & \shortstack[l]{$S(B,C,R):0.187/0.082/0.564\;(13)$ \\ $S(H,R):0.146/0.069/0.527\;(29)$ \\ $S(C,R):0.121/0.061/0.496\;(28)$} \\
 & $R$ & \shortstack[l]{$S(H,B,C,R):0.251/0.054/0.784\;(16)$ \\ $S(B,C,R):0.157/0.043/0.727\;(23)$ \\ $S(H,B,R):0.205/0.043/0.791\;(11)$} & \shortstack[l]{$S(B,C,R):0.215/0.070/0.675\;(16)$ \\ $S(H,R):0.128/0.054/0.581\;(21)$ \\ $S(B,R):0.135/0.052/0.614\;(19)$} \\
\bottomrule
\end{tabular}
\end{table*}

\begin{table*}
\centering
\scriptsize
\caption{Ranked synergistic components for the F5100 set $(F,B,C,R)$ with common lags. Notation and independent peak selection follow Table~\ref{tab:cross_u_total_wings}. The three entries in a cell need not share a lag and must not be added. No significance threshold is applied.}
\label{tab:synergistic_continuum_wings}
\setlength{\tabcolsep}{5pt}
\renewcommand{\arraystretch}{1.15}
\begin{tabular}{ccp{0.39\textwidth}p{0.39\textwidth}}
\toprule
Epoch & Target & Two bins: $a:f_a/g_a/\ell\;(L)$ & Three bins: $a:f_a/g_a/\ell\;(L)$ \\
\midrule
1989 & $F$ & \shortstack[l]{$S(F,B):0.229/0.086/0.625\;(15)$ \\ $S(F,B,C):0.197/0.071/0.642\;(25)$ \\ $S(F,C):0.206/0.063/0.694\;(17)$} & \shortstack[l]{$S(F,B,C,R):0.201/0.111/0.446\;(26)$ \\ $S(F,B,R):0.150/0.096/0.362\;(22)$ \\ $S(F,C):0.160/0.093/0.415\;(16)$} \\
 & $B$ & \shortstack[l]{$S(F,R):0.263/0.193/0.267\;(16)$ \\ $S(B,R):0.184/0.162/0.122\;(8)$ \\ $S(F,B,R):0.137/0.103/0.249\;(20)$} & \shortstack[l]{$S(F,B):0.209/0.170/0.188\;(30)$ \\ $S(F,B,R):0.111/0.104/0.066\;(8)$ \\ $S(F,R):0.117/0.091/0.223\;(17)$} \\
 & $C$ & \shortstack[l]{$S(F,C):0.178/0.124/0.304\;(13)$ \\ $S(B,R):0.150/0.114/0.239\;(8)$ \\ $S(F,R):0.153/0.107/0.304\;(13)$} & \shortstack[l]{$S(F,B):0.119/0.105/0.121\;(9)$ \\ $S(F,C):0.117/0.099/0.157\;(18)$ \\ $S(F,R):0.097/0.080/0.174\;(16)$} \\
 & $R$ & \shortstack[l]{$S(F,C):0.239/0.119/0.504\;(30)$ \\ $S(F,B):0.196/0.100/0.489\;(11)$ \\ $S(F,R):0.170/0.090/0.474\;(14)$} & \shortstack[l]{$S(F,B,R):0.142/0.109/0.230\;(24)$ \\ $S(F,B,C,R):0.141/0.090/0.365\;(28)$ \\ $S(F,R):0.129/0.084/0.350\;(10)$} \\
\addlinespace[3pt]
1990 & $F$ & \shortstack[l]{$S(F,B):0.414/0.148/0.642\;(20)$ \\ $S(F,C):0.207/0.073/0.650\;(28)$ \\ $S(F,R):0.140/0.049/0.652\;(27)$} & \shortstack[l]{$S(F,C):0.232/0.149/0.356\;(15)$ \\ $S(F,R):0.220/0.142/0.354\;(16)$ \\ $S(F,B):0.164/0.097/0.406\;(20)$} \\
 & $B$ & \shortstack[l]{$S(F,R):0.136/0.105/0.229\;(9)$ \\ $S(F,B):0.150/0.104/0.305\;(13)$ \\ $S(F,C):0.117/0.090/0.228\;(10)$} & \shortstack[l]{$S(F,B):0.304/0.224/0.263\;(12)$ \\ $S(F,C):0.156/0.114/0.264\;(11)$ \\ $S(F,R):0.148/0.102/0.306\;(25)$} \\
 & $C$ & \shortstack[l]{$S(F,B):0.250/0.150/0.399\;(12)$ \\ $S(F,C):0.217/0.128/0.409\;(13)$ \\ $S(F,B,C,R):0.068/0.048/0.292\;(7)$} & \shortstack[l]{$S(F,C):0.147/0.115/0.220\;(12)$ \\ $S(F,R):0.162/0.109/0.325\;(30)$ \\ $S(F,B):0.110/0.085/0.221\;(13)$} \\
 & $R$ & \shortstack[l]{$S(F,B):0.192/0.092/0.520\;(14)$ \\ $S(F,R):0.176/0.080/0.544\;(15)$ \\ $S(C,R):0.086/0.077/0.101\;(2)$} & \shortstack[l]{$S(F,R):0.254/0.160/0.371\;(27)$ \\ $S(F,B):0.213/0.148/0.305\;(10)$ \\ $S(F,C):0.130/0.080/0.388\;(19)$} \\
\addlinespace[3pt]
1991 & $F$ & \shortstack[l]{$S(B,R):0.158/0.052/0.672\;(28)$ \\ $S(F,C,R):0.178/0.050/0.718\;(11)$ \\ $S(C,R):0.136/0.039/0.713\;(29)$} & \shortstack[l]{$S(F,R):0.229/0.101/0.557\;(30)$ \\ $S(F,C,R):0.167/0.078/0.533\;(8)$ \\ $S(C,R):0.164/0.070/0.574\;(18)$} \\
 & $B$ & \shortstack[l]{$S(F,B):0.235/0.138/0.415\;(30)$ \\ $S(F,R):0.202/0.122/0.398\;(26)$ \\ $S(F,C):0.098/0.076/0.227\;(11)$} & \shortstack[l]{$S(B,R):0.200/0.140/0.297\;(29)$ \\ $S(C,R):0.121/0.085/0.297\;(28)$ \\ $S(F,C):0.118/0.083/0.295\;(18)$} \\
 & $C$ & \shortstack[l]{$S(F,B):0.205/0.187/0.087\;(20)$ \\ $S(F,R):0.136/0.125/0.087\;(20)$ \\ $S(F,C):0.083/0.073/0.127\;(15)$} & \shortstack[l]{$S(B,R):0.147/0.101/0.314\;(27)$ \\ $S(F,R):0.126/0.085/0.327\;(30)$ \\ $S(F,B,R):0.099/0.069/0.299\;(26)$} \\
 & $R$ & \shortstack[l]{$S(F,R):0.140/0.099/0.290\;(6)$ \\ $S(F,B):0.150/0.098/0.346\;(15)$ \\ $S(F,B,R):0.184/0.081/0.560\;(29)$} & \shortstack[l]{$S(F,R):0.188/0.127/0.327\;(10)$ \\ $S(F,C,R):0.199/0.122/0.388\;(19)$ \\ $S(B,R):0.221/0.117/0.470\;(27)$} \\
\addlinespace[3pt]
1992 & $F$ & \shortstack[l]{$S(F,B):0.341/0.064/0.812\;(26)$ \\ $S(F,R):0.342/0.055/0.840\;(30)$ \\ $S(F,B,R):0.052/0.026/0.502\;(7)$} & \shortstack[l]{$S(F,B):0.200/0.074/0.629\;(26)$ \\ $S(F,R):0.145/0.053/0.633\;(27)$ \\ $S(F,B,C):0.109/0.042/0.615\;(25)$} \\
 & $B$ & \shortstack[l]{$S(F,B):0.112/0.084/0.250\;(11)$ \\ $S(F,C,R):0.087/0.075/0.144\;(9)$ \\ $S(F,C):0.089/0.073/0.178\;(10)$} & \shortstack[l]{$S(F,C):0.106/0.075/0.293\;(13)$ \\ $S(B,R):0.057/0.049/0.132\;(5)$ \\ $S(F,B):0.079/0.048/0.397\;(22)$} \\
 & $C$ & \shortstack[l]{$S(F,B):0.159/0.121/0.239\;(7)$ \\ $S(F,R):0.107/0.083/0.227\;(6)$ \\ $S(F,C):0.176/0.063/0.643\;(24)$} & \shortstack[l]{$S(F,R):0.074/0.055/0.248\;(8)$ \\ $S(F,B):0.085/0.052/0.382\;(17)$ \\ $S(F,C):0.071/0.050/0.290\;(10)$} \\
 & $R$ & \shortstack[l]{$S(F,B):0.139/0.104/0.251\;(5)$ \\ $S(F,R):0.076/0.055/0.276\;(6)$ \\ $S(F,B,C):0.092/0.039/0.576\;(18)$} & \shortstack[l]{$S(F,B):0.202/0.090/0.557\;(29)$ \\ $S(F,R):0.164/0.077/0.530\;(28)$ \\ $S(F,C):0.180/0.076/0.576\;(30)$} \\
\addlinespace[3pt]
1993 & $F$ & \shortstack[l]{$S(F,B):0.185/0.086/0.532\;(13)$ \\ $S(F,B,C,R):0.330/0.075/0.772\;(28)$ \\ $S(F,R):0.123/0.061/0.501\;(10)$} & \shortstack[l]{$S(F,C):0.161/0.100/0.379\;(15)$ \\ $S(F,B,C,R):0.192/0.099/0.482\;(29)$ \\ $S(F,B):0.141/0.083/0.411\;(16)$} \\
 & $B$ & \shortstack[l]{$S(F,C):0.250/0.108/0.566\;(12)$ \\ $S(F,B,C,R):0.218/0.083/0.619\;(23)$ \\ $S(F,B,C):0.137/0.063/0.538\;(4)$} & \shortstack[l]{$S(F,B,C,R):0.122/0.087/0.289\;(6)$ \\ $S(F,B):0.111/0.076/0.319\;(12)$ \\ $S(F,B,R):0.112/0.076/0.323\;(14)$} \\
 & $C$ & \shortstack[l]{$S(F,R):0.401/0.172/0.570\;(28)$ \\ $S(F,C):0.234/0.111/0.524\;(9)$ \\ $S(F,B):0.125/0.059/0.529\;(25)$} & \shortstack[l]{$S(F,C):0.117/0.085/0.271\;(6)$ \\ $S(F,B):0.128/0.076/0.403\;(30)$ \\ $S(F,R):0.115/0.076/0.343\;(29)$} \\
 & $R$ & \shortstack[l]{$S(F,B,C,R):0.363/0.085/0.767\;(13)$ \\ $S(B,C,R):0.222/0.051/0.769\;(15)$ \\ $S(F,C,R):0.129/0.044/0.657\;(24)$} & \shortstack[l]{$S(B,C,R):0.217/0.093/0.572\;(8)$ \\ $S(F,B,C,R):0.188/0.092/0.513\;(11)$ \\ $S(F,R):0.144/0.066/0.540\;(12)$} \\
\bottomrule
\end{tabular}
\end{table*}

\paragraph*{Redundant influences and simpler predictor sets.}
Redundancy identifies which observations supply common information about a specified target. In the total set the largest shared terms frequently involve $H$ and several of its components. For the 1992 blue target, $R(H,B,C,R)$ reaches $g=0.681$ at 2~d with two states and $g=0.709$ at 3~d with three, with leakage 0.100 and 0.076. At the same 2-d lag the three-state value is still 0.673 with leakage 0.064. Shared illumination, overlapping responses and the construction $H\simeq B+C+R$ all contribute to this coherent information structure.

The F5100 set exposes more specific shared line histories. In 1991 the core target has $R(B,C,R)$ with $g=0.458$ at 4~d and almost zero leakage with two states, while the leading three-state pair is $R(B,C)$, reaching $g=0.403$ at 2~d with leakage 0.059. Tables~\ref{tab:redundant_total_wings} and \ref{tab:redundant_continuum_wings} identify the exact group rather than assigning an undifferentiated redundant budget to every line. A weak pair-only $R(B,R)$ also need not mean little shared wing information: that information can occur in $R(B,C,R)$ or $R(F,B,C,R)$.

The unequal-lag examples below reveal particularly large $R(C,R)$ and $R(B,C)$ pairs. Either member supplies the information represented by that shared term, making such pairs natural candidates for a reduced observational or predictive model. The omission criterion in equation~\eqref{eq:predictor_omission} determines whether the complete predictor can be dropped once its additional contributions are considered. This connects redundancy to a practical scientific use while retaining the distinct joint pathways found in the synergy tables.

\begin{table*}
\centering
\scriptsize
\caption{Ranked redundant components for the total set $(H,B,C,R)$ with common lags. Notation and independent peak selection follow Table~\ref{tab:cross_u_total_wings}. The three entries in a cell need not share a lag and must not be added. No significance threshold is applied.}
\label{tab:redundant_total_wings}
\setlength{\tabcolsep}{5pt}
\renewcommand{\arraystretch}{1.15}
\begin{tabular}{ccp{0.39\textwidth}p{0.39\textwidth}}
\toprule
Epoch & Target & Two bins: $a:f_a/g_a/\ell\;(L)$ & Three bins: $a:f_a/g_a/\ell\;(L)$ \\
\midrule
1989 & $H$ & \shortstack[l]{$R(H,B,C,R):0.511/0.471/0.078\;(2)$ \\ $R(H,C,R):0.191/0.179/0.065\;(1)$ \\ $R(H,B,C):0.108/0.051/0.528\;(11)$} & \shortstack[l]{$R(H,B,C,R):0.516/0.417/0.191\;(4)$ \\ $R(H,B,C):0.265/0.234/0.118\;(1)$ \\ $R(H,B):0.119/0.063/0.469\;(15)$} \\
 & $B$ & \shortstack[l]{$R(H,B,C,R):0.540/0.416/0.230\;(7)$ \\ $R(H,B,C):0.182/0.153/0.163\;(1)$ \\ $R(H,B):0.112/0.083/0.261\;(5)$} & \shortstack[l]{$R(H,B,C,R):0.501/0.388/0.226\;(6)$ \\ $R(H,B,C):0.334/0.307/0.079\;(1)$ \\ $R(H,B):0.119/0.103/0.134\;(3)$} \\
 & $C$ & \shortstack[l]{$R(H,B,C,R):0.553/0.485/0.123\;(2)$ \\ $R(H,B,C):0.109/0.101/0.072\;(1)$ \\ $R(B,C):0.065/0.048/0.251\;(4)$} & \shortstack[l]{$R(H,B,C,R):0.477/0.357/0.251\;(6)$ \\ $R(H,B,C):0.358/0.319/0.108\;(1)$ \\ $R(H,B):0.161/0.097/0.400\;(14)$} \\
 & $R$ & \shortstack[l]{$R(H,B,C,R):0.398/0.298/0.250\;(2)$ \\ $R(H,R):0.218/0.181/0.171\;(1)$ \\ $R(H,C,R):0.191/0.158/0.171\;(1)$} & \shortstack[l]{$R(H,B,C,R):0.450/0.297/0.340\;(3)$ \\ $R(H,B,C):0.270/0.128/0.523\;(15)$ \\ $R(H,C,R):0.093/0.076/0.185\;(1)$} \\
\addlinespace[3pt]
1990 & $H$ & \shortstack[l]{$R(H,B,C,R):0.658/0.619/0.059\;(1)$ \\ $R(H,C,R):0.176/0.169/0.039\;(3)$ \\ $R(H,R):0.038/0.036/0.064\;(2)$} & \shortstack[l]{$R(H,B,C,R):0.692/0.633/0.085\;(2)$ \\ $R(H,C,R):0.062/0.059/0.052\;(1)$ \\ $R(H,B,C):0.105/0.039/0.631\;(25)$} \\
 & $B$ & \shortstack[l]{$R(H,B,C,R):0.669/0.544/0.187\;(5)$ \\ $R(H,B):0.175/0.155/0.114\;(2)$ \\ $R(H,C,R):0.131/0.088/0.330\;(8)$} & \shortstack[l]{$R(H,B,C,R):0.644/0.470/0.271\;(3)$ \\ $R(H,B,C):0.121/0.105/0.131\;(1)$ \\ $R(H,B):0.066/0.057/0.131\;(1)$} \\
 & $C$ & \shortstack[l]{$R(H,B,C,R):0.544/0.492/0.095\;(1)$ \\ $R(H,C,R):0.161/0.146/0.095\;(1)$ \\ $R(H,C):0.097/0.083/0.140\;(2)$} & \shortstack[l]{$R(H,B,C,R):0.621/0.545/0.123\;(3)$ \\ $R(H,C):0.108/0.103/0.047\;(1)$ \\ $R(H,C,R):0.095/0.075/0.213\;(6)$} \\
 & $R$ & \shortstack[l]{$R(H,B,C,R):0.524/0.478/0.088\;(1)$ \\ $R(H,C,R):0.237/0.216/0.088\;(1)$ \\ $R(C,R):0.063/0.056/0.114\;(2)$} & \shortstack[l]{$R(H,B,C,R):0.680/0.509/0.251\;(3)$ \\ $R(H,R):0.135/0.116/0.141\;(1)$ \\ $R(H,C,R):0.058/0.049/0.141\;(1)$} \\
\addlinespace[3pt]
1991 & $H$ & \shortstack[l]{$R(H,B,C,R):0.715/0.638/0.107\;(3)$ \\ $R(H,B,C):0.158/0.148/0.062\;(1)$ \\ $R(C,R):0.087/0.071/0.187\;(6)$} & \shortstack[l]{$R(H,B,C):0.424/0.399/0.059\;(2)$ \\ $R(H,B,C,R):0.397/0.384/0.031\;(1)$ \\ $R(H,C):0.050/0.048/0.031\;(1)$} \\
 & $B$ & \shortstack[l]{$R(H,B,C,R):0.698/0.627/0.101\;(4)$ \\ $R(H,B,C):0.183/0.175/0.044\;(1)$ \\ $R(H,B):0.145/0.139/0.044\;(1)$} & \shortstack[l]{$R(H,B,C,R):0.439/0.375/0.145\;(6)$ \\ $R(H,B,C):0.383/0.371/0.032\;(1)$ \\ $R(H,B):0.078/0.075/0.032\;(1)$} \\
 & $C$ & \shortstack[l]{$R(H,B,C,R):0.611/0.594/0.028\;(4)$ \\ $R(H,B,C):0.227/0.223/0.019\;(1)$ \\ $R(H,C):0.109/0.107/0.019\;(1)$} & \shortstack[l]{$R(H,B,C):0.431/0.402/0.068\;(2)$ \\ $R(H,B,C,R):0.453/0.398/0.121\;(6)$ \\ $R(H,C):0.084/0.079/0.056\;(1)$} \\
 & $R$ & \shortstack[l]{$R(H,B,C,R):0.515/0.442/0.142\;(1)$ \\ $R(H,B,C):0.340/0.094/0.725\;(30)$ \\ $R(C,R):0.120/0.065/0.459\;(6)$} & \shortstack[l]{$R(H,B,C,R):0.593/0.330/0.443\;(5)$ \\ $R(H,B,C):0.244/0.112/0.539\;(14)$ \\ $R(H,B):0.091/0.041/0.554\;(8)$} \\
\addlinespace[3pt]
1992 & $H$ & \shortstack[l]{$R(H,B,C,R):0.709/0.679/0.042\;(1)$ \\ $R(H,C,R):0.104/0.080/0.229\;(4)$ \\ $R(H,R):0.069/0.066/0.042\;(1)$} & \shortstack[l]{$R(H,B,C,R):0.660/0.612/0.073\;(1)$ \\ $R(H,C,R):0.171/0.152/0.110\;(2)$ \\ $R(H,C):0.069/0.064/0.073\;(1)$} \\
 & $B$ & \shortstack[l]{$R(H,B,C,R):0.757/0.681/0.100\;(2)$ \\ $R(H,C,R):0.142/0.103/0.275\;(8)$ \\ $R(H,B):0.081/0.076/0.062\;(1)$} & \shortstack[l]{$R(H,B,C,R):0.768/0.709/0.076\;(3)$ \\ $R(H,B,C):0.077/0.073/0.050\;(1)$ \\ $R(H,C,R):0.066/0.047/0.294\;(9)$} \\
 & $C$ & \shortstack[l]{$R(H,B,C,R):0.666/0.611/0.083\;(1)$ \\ $R(H,C):0.104/0.096/0.083\;(1)$ \\ $R(H,C,R):0.101/0.085/0.159\;(2)$} & \shortstack[l]{$R(H,B,C,R):0.664/0.635/0.044\;(1)$ \\ $R(H,C,R):0.160/0.142/0.114\;(2)$ \\ $R(H,C):0.079/0.075/0.044\;(1)$} \\
 & $R$ & \shortstack[l]{$R(H,B,C,R):0.608/0.522/0.141\;(1)$ \\ $R(H,C,R):0.152/0.073/0.520\;(7)$ \\ $R(H,R):0.048/0.041/0.141\;(1)$} & \shortstack[l]{$R(H,B,C,R):0.608/0.544/0.105\;(1)$ \\ $R(H,C,R):0.115/0.103/0.105\;(1)$ \\ $R(H,R):0.065/0.030/0.537\;(11)$} \\
\addlinespace[3pt]
1993 & $H$ & \shortstack[l]{$R(H,C):0.398/0.158/0.602\;(5)$ \\ $R(H,B,C):0.189/0.144/0.235\;(1)$ \\ $R(H,B,C,R):0.170/0.130/0.235\;(1)$} & \shortstack[l]{$R(H,B,C,R):0.273/0.227/0.171\;(1)$ \\ $R(H,B,C):0.255/0.212/0.171\;(1)$ \\ $R(H,C):0.173/0.139/0.197\;(2)$} \\
 & $B$ & \shortstack[l]{$R(H,B,C):0.341/0.206/0.397\;(2)$ \\ $R(H,B):0.201/0.121/0.397\;(2)$ \\ $R(H,C):0.363/0.112/0.690\;(9)$} & \shortstack[l]{$R(H,B,C):0.424/0.309/0.272\;(2)$ \\ $R(H,B,C,R):0.181/0.120/0.338\;(5)$ \\ $R(H,C):0.131/0.068/0.482\;(15)$} \\
 & $C$ & \shortstack[l]{$R(H,B,C):0.184/0.147/0.202\;(1)$ \\ $R(H,C):0.179/0.143/0.202\;(1)$ \\ $R(H,B,C,R):0.066/0.053/0.202\;(1)$} & \shortstack[l]{$R(H,B,C):0.253/0.221/0.124\;(1)$ \\ $R(H,B,C,R):0.197/0.173/0.124\;(1)$ \\ $R(H,C):0.180/0.144/0.203\;(2)$} \\
 & $R$ & \shortstack[l]{$R(H,R):0.119/0.091/0.233\;(1)$ \\ $R(H,B,C,R):0.342/0.088/0.743\;(29)$ \\ $R(H,B,C):0.204/0.055/0.729\;(24)$} & \shortstack[l]{$R(H,B,C,R):0.360/0.152/0.578\;(30)$ \\ $R(H,C,R):0.104/0.073/0.292\;(1)$ \\ $R(H,B,C):0.173/0.070/0.592\;(24)$} \\
\bottomrule
\end{tabular}
\end{table*}

\begin{table*}
\centering
\scriptsize
\caption{Ranked redundant components for the F5100 set $(F,B,C,R)$ with common lags. Notation and independent peak selection follow Table~\ref{tab:cross_u_total_wings}. The three entries in a cell need not share a lag and must not be added. No significance threshold is applied.}
\label{tab:redundant_continuum_wings}
\setlength{\tabcolsep}{5pt}
\renewcommand{\arraystretch}{1.15}
\begin{tabular}{ccp{0.39\textwidth}p{0.39\textwidth}}
\toprule
Epoch & Target & Two bins: $a:f_a/g_a/\ell\;(L)$ & Three bins: $a:f_a/g_a/\ell\;(L)$ \\
\midrule
1989 & $F$ & \shortstack[l]{$R(B,C):0.236/0.082/0.651\;(30)$ \\ $R(F,B,C,R):0.076/0.060/0.218\;(1)$ \\ $R(F,B,C):0.131/0.040/0.694\;(17)$} & \shortstack[l]{$R(F,B,C,R):0.151/0.116/0.234\;(1)$ \\ $R(B,C,R):0.098/0.060/0.392\;(23)$ \\ $R(F,B,C):0.044/0.033/0.234\;(1)$} \\
 & $B$ & \shortstack[l]{$R(F,B,C,R):0.355/0.278/0.217\;(11)$ \\ $R(B,C,R):0.286/0.219/0.232\;(4)$ \\ $R(B,C):0.228/0.198/0.132\;(1)$} & \shortstack[l]{$R(F,B,C,R):0.363/0.321/0.115\;(9)$ \\ $R(B,C):0.333/0.307/0.076\;(1)$ \\ $R(B,C,R):0.111/0.103/0.076\;(1)$} \\
 & $C$ & \shortstack[l]{$R(B,C,R):0.382/0.359/0.061\;(1)$ \\ $R(F,B,C,R):0.397/0.295/0.259\;(9)$ \\ $R(B,C):0.142/0.116/0.180\;(4)$} & \shortstack[l]{$R(B,C):0.358/0.324/0.095\;(1)$ \\ $R(F,B,C,R):0.320/0.281/0.121\;(9)$ \\ $R(B,C,R):0.190/0.172/0.095\;(1)$} \\
 & $R$ & \shortstack[l]{$R(F,B,C,R):0.388/0.206/0.468\;(10)$ \\ $R(B,C,R):0.259/0.198/0.234\;(2)$ \\ $R(C,R):0.192/0.158/0.174\;(1)$} & \shortstack[l]{$R(F,B,C,R):0.351/0.215/0.388\;(8)$ \\ $R(B,C,R):0.180/0.129/0.283\;(3)$ \\ $R(F,B,C):0.139/0.092/0.342\;(13)$} \\
\addlinespace[3pt]
1990 & $F$ & \shortstack[l]{$R(B,C,R):0.244/0.090/0.633\;(30)$ \\ $R(C,R):0.227/0.083/0.633\;(29)$ \\ $R(F,B):0.086/0.067/0.224\;(1)$} & \shortstack[l]{$R(C,R):0.192/0.101/0.471\;(30)$ \\ $R(F,B,C,R):0.146/0.091/0.381\;(19)$ \\ $R(F,C):0.077/0.050/0.354\;(16)$} \\
 & $B$ & \shortstack[l]{$R(B,C,R):0.361/0.339/0.060\;(1)$ \\ $R(F,B,C,R):0.419/0.323/0.228\;(10)$ \\ $R(C,R):0.116/0.092/0.204\;(8)$} & \shortstack[l]{$R(F,B,C,R):0.447/0.329/0.264\;(11)$ \\ $R(B,C,R):0.340/0.312/0.082\;(1)$ \\ $R(B,C):0.115/0.105/0.082\;(1)$} \\
 & $C$ & \shortstack[l]{$R(B,C,R):0.443/0.408/0.079\;(1)$ \\ $R(F,B,C,R):0.386/0.245/0.364\;(10)$ \\ $R(C,R):0.158/0.146/0.079\;(1)$} & \shortstack[l]{$R(B,C,R):0.467/0.454/0.027\;(1)$ \\ $R(F,B,C,R):0.372/0.292/0.215\;(10)$ \\ $R(C,R):0.104/0.084/0.191\;(8)$} \\
 & $R$ & \shortstack[l]{$R(B,C,R):0.439/0.398/0.094\;(1)$ \\ $R(C,R):0.298/0.268/0.101\;(2)$ \\ $R(F,B,C,R):0.433/0.240/0.447\;(11)$} & \shortstack[l]{$R(B,C,R):0.441/0.389/0.118\;(1)$ \\ $R(F,B,C,R):0.393/0.291/0.260\;(8)$ \\ $R(F,B,C):0.114/0.062/0.459\;(30)$} \\
\addlinespace[3pt]
1991 & $F$ & \shortstack[l]{$R(B,R):0.241/0.071/0.707\;(30)$ \\ $R(B,C,R):0.206/0.060/0.707\;(30)$ \\ $R(F,B,C,R):0.084/0.059/0.295\;(1)$} & \shortstack[l]{$R(F,B,C):0.141/0.066/0.533\;(8)$ \\ $R(F,B):0.077/0.064/0.174\;(1)$ \\ $R(F,B,C,R):0.073/0.060/0.174\;(1)$} \\
 & $B$ & \shortstack[l]{$R(B,C,R):0.559/0.515/0.079\;(4)$ \\ $R(F,B,C,R):0.344/0.234/0.320\;(18)$ \\ $R(B,C):0.181/0.175/0.035\;(1)$} & \shortstack[l]{$R(B,C):0.356/0.346/0.030\;(1)$ \\ $R(F,B,C,R):0.316/0.258/0.181\;(13)$ \\ $R(B,C,R):0.148/0.144/0.030\;(1)$} \\
 & $C$ & \shortstack[l]{$R(B,C,R):0.458/0.458/{<}0.001\;(4)$ \\ $R(F,B,C,R):0.386/0.347/0.101\;(19)$ \\ $R(B,C):0.229/0.223/0.025\;(1)$} & \shortstack[l]{$R(B,C):0.429/0.403/0.059\;(2)$ \\ $R(F,B,C,R):0.322/0.286/0.113\;(9)$ \\ $R(B,C,R):0.219/0.206/0.059\;(2)$} \\
 & $R$ & \shortstack[l]{$R(B,C,R):0.406/0.357/0.120\;(1)$ \\ $R(F,B,C,R):0.339/0.241/0.290\;(6)$ \\ $R(B,C):0.325/0.137/0.579\;(30)$} & \shortstack[l]{$R(F,B,C,R):0.384/0.265/0.311\;(6)$ \\ $R(B,C,R):0.271/0.231/0.146\;(1)$ \\ $R(F,B):0.229/0.156/0.319\;(11)$} \\
\addlinespace[3pt]
1992 & $F$ & \shortstack[l]{$R(F,B,C,R):0.288/0.208/0.277\;(1)$ \\ $R(F,C,R):0.113/0.073/0.349\;(2)$ \\ $R(F,R):0.094/0.059/0.372\;(4)$} & \shortstack[l]{$R(F,B,C,R):0.398/0.319/0.200\;(1)$ \\ $R(F,R):0.124/0.099/0.200\;(1)$ \\ $R(C,R):0.132/0.059/0.548\;(19)$} \\
 & $B$ & \shortstack[l]{$R(F,B,C,R):0.517/0.437/0.155\;(7)$ \\ $R(B,C,R):0.449/0.422/0.061\;(1)$ \\ $R(F,C,R):0.158/0.130/0.178\;(10)$} & \shortstack[l]{$R(F,B,C,R):0.610/0.481/0.211\;(8)$ \\ $R(B,C,R):0.354/0.325/0.084\;(3)$ \\ $R(F,R):0.184/0.124/0.324\;(16)$} \\
 & $C$ & \shortstack[l]{$R(F,B,C,R):0.545/0.428/0.214\;(4)$ \\ $R(B,C,R):0.292/0.272/0.070\;(1)$ \\ $R(C,R):0.142/0.118/0.163\;(3)$} & \shortstack[l]{$R(F,B,C,R):0.530/0.454/0.143\;(3)$ \\ $R(B,C,R):0.233/0.222/0.050\;(1)$ \\ $R(C,R):0.160/0.142/0.114\;(2)$} \\
 & $R$ & \shortstack[l]{$R(F,B,C,R):0.536/0.409/0.237\;(3)$ \\ $R(B,C,R):0.160/0.141/0.114\;(1)$ \\ $R(F,C,R):0.144/0.104/0.275\;(7)$} & \shortstack[l]{$R(F,B,C,R):0.552/0.515/0.067\;(1)$ \\ $R(F,R):0.186/0.144/0.226\;(6)$ \\ $R(C,R):0.086/0.080/0.067\;(1)$} \\
\addlinespace[3pt]
1993 & $F$ & \shortstack[l]{$R(C,R):0.311/0.062/0.799\;(27)$ \\ $R(F,R):0.154/0.055/0.642\;(17)$ \\ $R(B,C,R):0.156/0.037/0.766\;(25)$} & \shortstack[l]{$R(F,B,C,R):0.130/0.077/0.403\;(18)$ \\ $R(F,C):0.145/0.075/0.482\;(29)$ \\ $R(B,C):0.074/0.045/0.388\;(21)$} \\
 & $B$ & \shortstack[l]{$R(B,C):0.357/0.204/0.428\;(2)$ \\ $R(F,C):0.326/0.141/0.566\;(12)$ \\ $R(F,B,C):0.192/0.064/0.668\;(7)$} & \shortstack[l]{$R(B,C):0.393/0.301/0.234\;(2)$ \\ $R(F,C):0.186/0.129/0.307\;(13)$ \\ $R(F,B,C):0.153/0.108/0.299\;(10)$} \\
 & $C$ & \shortstack[l]{$R(F,R):0.337/0.145/0.570\;(28)$ \\ $R(B,C):0.156/0.126/0.191\;(1)$ \\ $R(F,C):0.246/0.125/0.492\;(10)$} & \shortstack[l]{$R(B,C):0.206/0.187/0.091\;(1)$ \\ $R(F,B,C,R):0.201/0.145/0.279\;(5)$ \\ $R(F,C):0.166/0.118/0.290\;(9)$} \\
 & $R$ & \shortstack[l]{$R(B,C,R):0.177/0.059/0.667\;(26)$ \\ $R(B,C):0.169/0.058/0.657\;(24)$ \\ $R(F,B,C,R):0.141/0.040/0.716\;(29)$} & \shortstack[l]{$R(B,C,R):0.258/0.112/0.568\;(28)$ \\ $R(B,C):0.154/0.075/0.513\;(24)$ \\ $R(C,R):0.074/0.054/0.264\;(1)$} \\
\bottomrule
\end{tabular}
\end{table*}

\subsection{Unequal lags: changes in specific predictor combinations}
\label{sec:mixed}
Table~\ref{tab:selected_mixed} gives selected named-component maxima for the two imposed wing offsets, with both bin counts and simultaneous leakage. The supplementary catalogue gives the top three components of every family for all 160 unequal-lag target--bin scans, using the same ranking as the common-lag tables.

The 1989 continuum--core influence persists under either offset. Blue-leading and red-leading scans both reach $g_{U(F)}=0.872$ at base lag 24~d with two states, with $f=0.949/0.918$ and leakage $0.080/0.050$, respectively. Since the continuum itself is unshifted, this remains a 24-d continuum history; the shifted wing is sampled at 29~d. These cases strengthen the identification of the continuum--core channel within different history arrangements.

Figure~\ref{fig:ep_blue_leading_1992_continuum_wings_core} shows the full 1992 core scan with the blue-wing history five days further back. Its continuum-unique $g$ maxima are 0.209 at 14~d with two states and 0.176 at 12~d with three, with leakage 0.381 and 0.236. They remain close to the common-lag values, 0.209 and 0.170 at the same respective lags. The two-state $S(F,B)$ relative fraction rises to 0.305 at base lag 30~d, where $g=0.091$ and leakage is 0.703. This identifies the joint continuum--older-blue-history contribution shown in the synergy panel.

\begin{SURDfigure}
\centering
\includegraphics[width=0.98\textwidth]{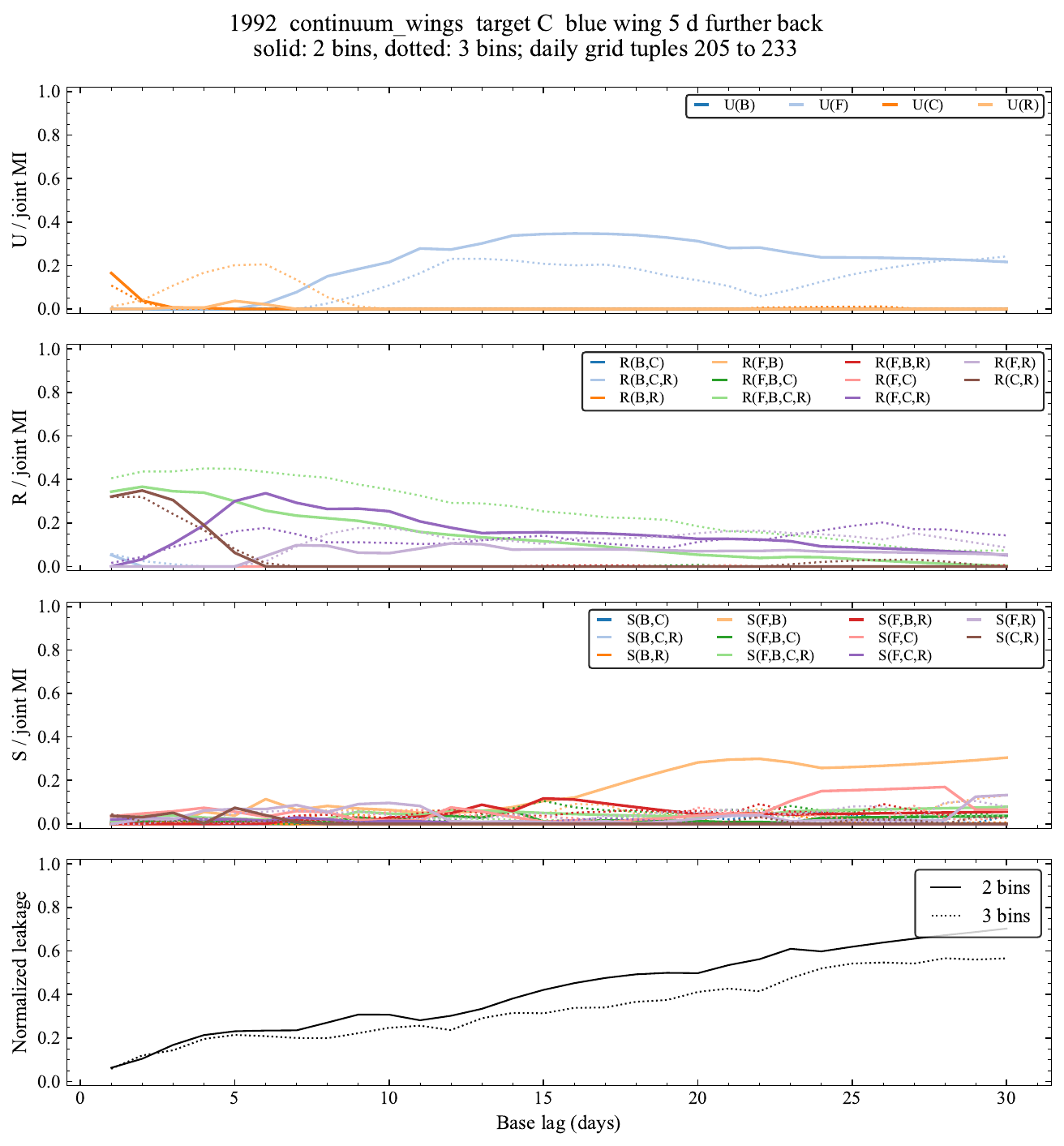}
\caption{Campaign 1992, F5100 set $(F,B,C,R)$, core target $C$, with the blue-wing predictor sampled five days further back: $X_B=B(t-L-5~\mathrm{d})$, while the other predictors are sampled at $t-L$. The U, R and S panels show all 26 fractions $f_a=I_a/I(Y;\mathbf{X})$; the bottom panel shows $\ell=H(Y\mid\mathbf{X})/H(Y)$. Solid curves use two flux bins and dotted curves three bins; the daily grid supplies 205--233 tuples per base lag. This imposed offset is a sensitivity configuration. Source: blue-leading atlas, page 27.}
\label{fig:ep_blue_leading_1992_continuum_wings_core}
\end{SURDfigure}

The shifted histories also resolve strong pairwise redundancy. For the 1989 core in the F5100 set, the red-leading configuration at base lag 1~d gives $R(B,C)$ with $g=0.319/0.381$ and leakage $0.052/0.079$ for two/three states. For the 1992 core with blue-leading histories, $R(C,R)$ contributes $g=0.313/0.282$ at base lag 2~d, with leakage $0.104/0.119$. The shared core--red contribution persists in both discretizations. Another instance is the 1990 core, blue-leading at 5~d, where $R(C,R)$ has $g=0.336/0.217$ and leakage $0.194/0.101$. These examples identify specific candidate input pairs for simplification; their simultaneous unique and synergistic terms remain available in the catalogue for assessing what information both inputs add.

Offset sensitivity itself is useful for identifying a joint history. For example, the 1991 red-leading core has a two-state $S(F,R)$ relative maximum $f=0.306$ at base lag 25~d, with $g=0.186$ and leakage 0.391. This combines the continuum at 25~d with the red wing at 30~d. Its three-state relative maximum is much smaller, with $g=0.030$, so this particular amplitude is sensitive to the flux resolution. The distinguished common-lag $S(F,B)$ and $S(F,R)$ terms described above provide the complementary reference.

At the same base lag of 20~d for this 1991 core target, the two-state $S(F,B)$ and $S(F,R)$ contributions are $0.187/0.125$ with common histories, $0.128/0.140$ with blue-leading histories, and $0.080/0.252$ with red-leading histories. Their simultaneous leakages are 0.087, 0.100 and 0.107, respectively. This directly identifies a change between the two named joint contributions while $U(F)$ remains zero. The red-leading $S(F,R)$ value of 0.252 is its largest $g$, distinct from the later maximum of $f$ quoted above. Common and shifted cases retain 138 and 121 tuples, respectively.

The 1992 $U(R)\to B$ channel has a different alignment dependence. It remains comparable under blue-leading histories, with maxima $g=0.130/0.170$ in the F5100 set and $0.109/0.153$ in the total set for two/three states. Under red-leading histories it is zero throughout both two-state scans and reaches only $0.012/0.014$ in the three-state F5100/total scans. This localizes the distinguished unique contribution to the common and blue-shifted history arrangements; the five-day older red history supplies a different information allocation.

If the shifted wing is the target, the target's own history also changes. For the 1993 F5100-set red target at base lag 1~d, changing to red-leading histories moves self-history from 1 to 6~d: $g_{U(R)}$ changes from $0.667/0.473$ to $0.185/0.049$, and leakage from $0.238/0.264$ to $0.602/0.465$. These configurations therefore probe the entire history vector. The imposed offset is a sampling choice; the source--target lag of a shifted variable is $L+5$~d, and the offset itself is not a measured propagation time.

\begin{SURDtable}
\centering
\caption{Selected named components with unequal predictor lags. Each entry is $(L;f,g,\ell)$ at that component's own maximum $g$, separately for each state count. H and F denote the total and F5100 sets. B+5 or R+5 samples the indicated wing at $L+5$ days while other histories are sampled at $L$. A dash denotes a component that is zero throughout the scan. The supplementary catalogue covers every target and all leading components.}
\label{tab:selected_mixed}
\small
\begin{tabular}{cccllll}
\toprule
Epoch & Set & Target & Offset & Component & Two bins: $(L;f,g,\ell)$ & Three bins: $(L;f,g,\ell)$ \\
\midrule
1989 & F & $C$ & $B+5$ & $U(F)$ & $(24;0.949,0.872,0.080)$ & $(25;0.477,0.345,0.277)$ \\
1989 & F & $C$ & $R+5$ & $U(F)$ & $(24;0.918,0.872,0.050)$ & $(26;0.433,0.311,0.282)$ \\
1989 & F & $C$ & $R+5$ & $R(B,C)$ & $(1;0.337,0.319,0.052)$ & $(1;0.413,0.381,0.079)$ \\
1990 & F & $C$ & $B+5$ & $R(C,R)$ & $(5;0.417,0.336,0.194)$ & $(5;0.241,0.217,0.101)$ \\
1990 & F & $B$ & $B+5$ & $S(F,B)$ & $(13;0.090,0.062,0.305)$ & $(12;0.225,0.172,0.236)$ \\
1991 & F & $C$ & $R+5$ & $S(F,R)$ & $(20;0.283,0.252,0.107)$ & $(16;0.043,0.032,0.255)$ \\
1991 & H & $B$ & $B+5$ & $S(C,R)$ & $(6;0.063,0.054,0.132)$ & $(26;0.227,0.158,0.305)$ \\
1992 & F & $B$ & $B+5$ & $U(R)$ & $(5;0.151,0.130,0.140)$ & $(9;0.212,0.170,0.199)$ \\
1992 & F & $B$ & $R+5$ & $U(R)$ & -- & $(4;0.014,0.012,0.124)$ \\
1992 & H & $B$ & $B+5$ & $U(R)$ & $(10;0.167,0.109,0.346)$ & $(9;0.211,0.153,0.273)$ \\
1992 & H & $B$ & $R+5$ & $U(R)$ & -- & $(24;0.047,0.014,0.692)$ \\
1992 & F & $C$ & $B+5$ & $R(C,R)$ & $(2;0.349,0.313,0.104)$ & $(1;0.320,0.301,0.058)$ \\
1993 & F & $C$ & $B+5$ & $S(F,R)$ & $(28;0.385,0.168,0.564)$ & $(25;0.106,0.069,0.344)$ \\
1993 & H & $B$ & $R+5$ & $S(B,C,R)$ & $(17;0.124,0.041,0.672)$ & $(24;0.188,0.085,0.550)$ \\
\bottomrule
\end{tabular}
\end{SURDtable}

\subsection{Comparison with Lu et al. and Xi et al.}
\label{sec:comparison}
SURD and reverberation mapping answer complementary questions. The analyses of \citet{Lu2022} and \citet{Xi2025} measure continuum-referenced line delays and their velocity structure, thereby constraining the response-weighted size and kinematics of the BLR. Our analysis identifies possible unique, shared and joint causal influences on each future light-curve state, together with the unresolved information measured by leakage. For example, $U(R)\to B$ identifies information specific to the red-wing history, $R(B,R)\to C$ identifies shared wing information about the core, and $S(F,R)\to C$ identifies a joint continuum--red-wing contribution. These distinctions add a multivariate causal-discovery perspective to the reverberation picture.

The observables must be matched when comparing time-scales. The published integrated H$\beta$ delays concern $F\to H$. Our total set contains $H$ but no $F$, while the F5100 set contains $F$ but no $H$; consequently the archived design contains no direct $F\to H$ pair. Its $U(F)\to B,C,R$ peaks describe the individual line bands. Table~\ref{tab:ep_lu} therefore compares characteristic component time-scales with the historical integrated delays compiled by \citet{Lu2022}, rather than treating their seasonal median as an integrated-line lag estimator. We use $L_{\rm rest}=L/(1+z)$ with $z=0.01717$. The maximum of $f_a$ describes the most prominent fractional allocation, whereas the maximum of $g_a$ describes the largest resolved contribution to target entropy; neither is an ICCF centroid.

The most repeatable continuum-referenced wing ordering occurs in 1989 and 1992. Let $D_f=L_{f,R}-L_{f,B}$, so a negative value denotes an earlier continuum-unique peak for the red target. In observed days, $D_f=-7/-6$ for two/three states in 1989 and $-5/-8$ in 1992. Using $g$ maxima gives $D_g=-4/-8$ and $-6/-8$, respectively. Thus both epochs retain red-before-blue information-peak ordering under both state counts and both normalizations. In 1990 the two-state blue-before-red ordering becomes a tie with three states; the three-state 1991 blue continuum-unique term is very small, and the two-state 1993 red term vanishes. Season-resolved analysis exposes these changes directly.

The directed wing calculation adds information beyond this ordering. In 1989, $U(R)\to B$ is zero in both sets and both state counts despite the earlier red continuum-associated peak. In 1992, by contrast, the nonzero $U(R)\to B$ values in Table~\ref{tab:directed_1992}, accompanied by zero reverse unique terms, identify an additional asymmetric candidate influence. The red history also supplies unique information about future $H$. A delayed common driver can therefore produce ordered responses without a separate unique inter-wing channel, while another epoch can exhibit both. The unequal-lag results specify the history alignment in which the 1992 unique asymmetry is expressed.

\begin{SURDtable}
\centering
\caption{Same-season comparison with the historical total-H$\beta$ ICCF lags compiled by \citet{Lu2022}, their Table~6. Published lags and median SURD peak lags are in rest-frame days. SURD medians use every strictly nonzero common-lag unique-continuum $f$ maximum across the line targets. The two right columns show red-minus-blue peak differences in \emph{observed} days for two/three bins, using $f$ and $g$ maxima; divide by 1.01717 for rest-frame differences. ``Weak'' flags the very small 1991 three-bin blue amplitude, not an applied significance cut. A dash means that a wing's unique-continuum peak is undefined.}
\label{tab:ep_lu}
\small
\begin{tabular}{ccrrll}
\toprule
& Published $\tau_{\rm H\beta}$ & \multicolumn{2}{c}{Median SURD $L_f/(1+z)$} & \multicolumn{2}{c}{Wing difference (2/3 bins)} \\ \cmidrule(lr){3-4}\cmidrule(lr){5-6} Epoch & Rest days & Two bins & Three bins & $L_{f,R}-L_{f,B}$ & $L_{g,R}-L_{g,B}$ \\
\midrule
1989 & $19.7^{+1.5}_{-1.5}$ & 22.61 & 24.58 & $-7/-6$ & $-4/-8$ \\
1990 & $18.6^{+2.1}_{-2.3}$ & 26.54 & 18.68 & $+6/0$ & $+5/0$ \\
1991 & $15.9^{+2.9}_{-2.5}$ & 20.15 & 19.66 & $-5/-8$ (weak) & $-5/-8$ (weak) \\
1992 & $11.0^{+1.9}_{-2.0}$ & 16.71 & 18.68 & $-5/-8$ & $-6/-8$ \\
1993 & $13.0^{+1.6}_{-1.4}$ & 22.61 & 19.66 & -- / $-15$ & -- / $-15$ \\
\bottomrule
\end{tabular}
\end{SURDtable}

Table~\ref{tab:ep_xi} summarizes the modern reverberation context. The 2019 and 2021 red-leading tendencies discussed by \citet{Lu2022} and \citet{Xi2025}, although the red- and blue-wing lags remain comparable, motivate an inflow-compatible interpretation of a red-before-blue response, possibly combined with virialized motion. Xi's 2023 blue-leading profile instead permits an outflow/virial mixture or anisotropic illumination associated with a variable disc wind. These alternatives suggest specific future SURD comparisons of $U(R)\to B$ and $U(B)\to R$, together with their redundant and synergistic counterparts. The modern campaigns and our 1989--1993 seasons are different data sets, so their classifications provide physical context rather than a transfer of the same seasonal geometry.

For the velocity analysis, Lu used uniformly spaced velocity intervals and Xi used 15 or 30 equal-flux intervals. Our fixed historical wavelength bands differ from both, and our two/three bins are flux-amplitude states. Xi does not tabulate a universal numerical lag for our coarse $B,C,R$ bands. In particular, the 2015 M-shaped profile has off-centre maxima and a central depression; it does not imply that every coarse core band follows both wings. The 2020 outflow-like profile is tentative in Xi's updated analysis, whereas Lu obtained no credible velocity-resolved profile for that season. These details prevent imposing a predetermined SURD component or leakage level from a qualitative velocity profile.

\begin{SURDtable}
\centering
\caption{Modern H$\beta$ reverberation lags and qualitative profile descriptions from \citet{Xi2025}, their Table~3 and Section~5.2; the pre-2023 integrated-line lags are compiled from \citet{Lu2022}. Lags are rest-frame ICCF centroids with published uncertainties. Profile descriptions are qualified interpretations of different observing seasons, not SURD classifications.}
\label{tab:ep_xi}
\small
\begin{tabular}{rcl}
\toprule
Season & $\tau_{\rm H\beta}$ (rest days) & Velocity-resolved profile interpretation \\
\midrule
2015 & $7.2^{+1.3}_{-0.4}$ & M-shaped; disc-like, with anisotropic illumination \\
2018 & $7.0^{+2.3}_{-3.4}$ & H$\beta$ less structured; no clear simple ordering \\
2019 & $8.9^{+2.0}_{-1.1}$ & General red-before-blue trend; mixed infall/virial \\
2020 & $10.0^{+3.3}_{-3.3}$ & Tentative outflow-like profile in Xi's reanalysis \\
2021 & $9.0^{+1.9}_{-2.5}$ & General red-before-blue trend; mixed infall/virial \\
2023 & $15.6^{+2.6}_{-2.9}$ & Blue-before-red; mixed outflow/virial possible \\
\bottomrule
\end{tabular}
\end{SURDtable}

A direct quantitative comparison would integrate the same spectra over matched velocity intervals and measure $\Delta\tau_{R-B}=\tau_R-\tau_B$, with negative values meaning that the red response precedes the blue. This signed difference of continuum-referenced delays is distinct from a source--target SURD peak and from the imposed five-day history offset. Examining both target directions, as here, allows possible directional asymmetry to be assessed while keeping all predictors earlier than their targets. SURD then extends the comparison by asking whether the earlier signal supplies unique information, duplicates another signal, or becomes informative jointly with it, and how much information remains unresolved.

\subsection{Astrophysical interpretation of the named influences}
\label{sec:astrophysics}
Photoionization by a variable central source provides a natural physical setting for the identified components \citep{Peterson2004ApJ,Lu2022,Xi2025}. Schematically, the line-band fluctuations can be written as
\begin{equation}
\delta Q_j(t)=\int_0^\infty\Psi_j(\tau)\,
\delta F_{\rm ion}(t-\tau)\,{\rm d}\tau+\epsilon_j(t),
\qquad j\in\{B,C,R\},
\label{eq:ep_response_interpretation}
\end{equation}
where $F_{\rm ion}$ is the unmeasured ionizing continuum, $\Psi_j$ describes a linearized response over delay, and $\epsilon_j$ collects additional variability. This is an interpretive model rather than a transfer function fitted by SURD. The measured optical $F$ traces the driver imperfectly, and each broad velocity interval combines gas with a range of response times.

The strong 1989 $U(F)\to C$ component identifies a candidate continuum-driven core response carrying information not supplied by any other observed history individually. Its persistence with two and three flux states, and with either wing offset, makes this a particularly useful physical pathway to investigate. The 1992 $U(R)\to B$ and $U(R)\to H$ components identify a different aspect of the response: the red-wing history is especially informative about subsequent blue-wing and integrated-line states. Together with the continuum-associated red-before-blue peak ordering, this is compatible with an inflow-weighted response within a mixed BLR, while a common driver and asymmetric illumination remain possible explanations. The information arrow connects sampled histories; it need not represent material or radiation passing directly from one wing-emitting region to the other.

The named synergies identify the combinations needed to recover additional information. The 1989 $S(F,R)\to B$, the 1990 $S(F,B)\to C$ and $S(F,C)\to C$, and the 1993 $S(F,R)\to C$ point to joint constraints involving the continuum and particular parts of the line profile. In 1991, $S(F,B)\to C$ and $S(F,R)\to C$ remain appreciable when the two-state $U(F)\to C$ is zero. Thus the absence of a unique continuum term can describe a change in how continuum information is expressed, rather than absence of a continuum-related influence. A line history may encode the delayed illumination and response state needed to interpret the optical continuum jointly. Such joint information can arise through response mixing and coarse flux states as well as nonlinear physical variability; SURD identifies the informative combination without requiring one specific gas-dynamical mechanism.

Redundancy describes the complementary shared part of the response. Strong $R(B,C)$ or $R(C,R)$ is consistent with common illumination and overlapping response functions. Either member can supply the shared contribution, suggesting useful reductions in the predictor set or priorities when choosing which light curves to monitor. Their unique and synergistic contributions indicate where retaining both may add information. In the total set, the large $R(H,B,C,R)$ additionally reflects the construction of the integrated line from its components and their shared spectral calibration. This distinguishes physically plausible coherent response from an independent interaction attributed to each member of a redundant group.

The astrophysical value of a component does not require it to approach unity. A noisy, spatially extended BLR can distribute information among several histories while leaving a large unresolved part. The recurring named pairs and lag-dependent unique channels therefore provide specific candidate pathways and observational tests even at moderate $g_a$. Conversely, a line history that predicts future optical $F$ can trace delayed or persistent information about the same central source; it should not automatically be interpreted as feedback from the line-emitting gas to the continuum source.

\subsection{Leakage and unresolved drivers across epochs}
\label{sec:sensitivity}
Causal leakage identifies how incomplete the specified history vector is for each target. Table~\ref{tab:leakage_cases} pairs selected low- and high-leakage cases with the largest individual synergy at that exact lag. The highest two-state leakage is 0.967 for the 1992 red target in the total set, with red-leading histories at base lag 30~d. Only 0.033 of the target entropy is resolved there. Its leading synergy is $S(B,R)$, with $f=0.291$ but $g=0.010$. With three states at the same lag, leakage is 0.797 and the leading synergy is $S(H,B)$ with $g=0.021$. The largest three-state leakage, 0.811, occurs for the 1992 core in the same set, configuration and base lag. These late-time targets are clear priorities for investigating additional driving information.

Substantial leakage also coexists with distinguished components. The 1993 total-set core's $U(R)$ contributes $g=0.128$ at 27~d even with $\ell=0.797$. In the F5100 set, its $S(F,R)$ peak has $g=0.172$ with $\ell=0.570$ at 28~d with two states, and $g=0.076$ with $\ell=0.343$ at 29~d with three. These cases identify a resolved unique or joint pathway and, separately, a sizable information deficit. The latter motivates observing the ionizing/UV continuum, testing variable obscuration or illumination, and representing a broader range of past response times.

Small leakage identifies where the observed states provide a more complete description. For the 1989 core, the two-state continuum-unique maximum at 24~d has leakage 0.047. For the 1991 core in the F5100 set, leakage at common lag 4~d is approximately zero with two states and 0.079 with three. The near-zero value means that the occupied predictor cells determine the binary target in the retained sample. Together with the 1991 joint-component examples, it locates epochs and histories in which much of the observed core variability is represented by the available variables.

\begin{SURDtable}
\centering
\caption{Leakage at selected fixed target--lag configurations, paired with the largest individual synergy at that same point. Both state counts use the displayed base lag. The synergy entries are $a:f_a/g_a$; these are not maxima over lag. F and H identify the F5100 and total sets. A leakage printed as 0.000 is zero to the displayed precision.}
\label{tab:leakage_cases}
\small
\begin{tabular}{ccclrrlrl}
\toprule
& & & & & \multicolumn{2}{c}{Two bins} & \multicolumn{2}{c}{Three bins} \\ \cmidrule(lr){6-7}\cmidrule(lr){8-9} Epoch & Set & Target & Configuration & $L$ & $\ell$ & Leading $S$: $f/g$ & $\ell$ & Leading $S$: $f/g$ \\
\midrule
1989 & F & $C$ & Common & 24 & 0.047 & $S(F,B,C):0.087/0.083$ & 0.284 & $S(F,B,C,R):0.068/0.049$ \\
1991 & F & $C$ & Common & 4 & 0.000 & $S(C,R):0.059/0.059$ & 0.079 & $S(B,C):0.029/0.026$ \\
1991 & F & $C$ & Common & 20 & 0.087 & $S(F,B):0.205/0.187$ & 0.316 & $S(F,C):0.077/0.053$ \\
1992 & H & $R$ & $R+5$ & 30 & 0.967 & $S(B,R):0.291/0.010$ & 0.797 & $S(H,B):0.105/0.021$ \\
1992 & H & $C$ & $R+5$ & 30 & 0.930 & $S(H,C,R):0.230/0.016$ & 0.811 & $S(H,B,C,R):0.116/0.022$ \\
1993 & H & $C$ & Common & 27 & 0.797 & $S(B,R):0.103/0.021$ & 0.517 & $S(H,R):0.117/0.057$ \\
\bottomrule
\end{tabular}
\end{SURDtable}

For common lags, the mean leakage over both sets and their targets is 0.49, 0.53, 0.43, 0.58 and 0.65 in 1989--1993 with two states. Table~\ref{tab:leakage_means} gives the separate set and configuration means. The 1993 scans have the largest mean unresolved fraction, whereas the 1991 scans have the smallest. Comparing the same line targets, replacing $H$ by $F$ reduces lag-averaged leakage in all 90 epoch--target--configuration--bin comparisons; the common-lag mean reductions are 0.205 and 0.160 for two and three states. The specific continuum-related components in the main tables identify where the added observed information appears.

These leakage patterns guide the search for hidden drivers, but do not identify a unique missing physical variable: noise, measurement uncertainty, omitted history and estimation effects also contribute to conditional entropy. Seasonal flux thresholds and sample occupancy change between the two discretizations. A less structured reverberation profile likewise does not by itself require high leakage. Physical interpretations can be tested by asking which additional measured driver or history reduces the unresolved fraction on matched samples.

\begin{SURDtable}
\centering
\caption{Mean normalized leakage over all 30 lags and four targets of each set. Each configuration lists two-bin and three-bin values. Because the fourth target differs between the sets, these means characterize each analysis separately; The text also compares the two sets with the line targets held fixed.}
\label{tab:leakage_means}
\small
\begin{tabular}{clrrrrrr}
\toprule
& & \multicolumn{2}{c}{Common} & \multicolumn{2}{c}{Blue leading} & \multicolumn{2}{c}{Red leading} \\ \cmidrule(lr){3-4}\cmidrule(lr){5-6}\cmidrule(lr){7-8} Epoch & Set & Two bins & Three bins & Two bins & Three bins & Two bins & Three bins \\
\midrule
1989 & Total & 0.607 & 0.446 & 0.616 & 0.437 & 0.605 & 0.451 \\
1989 & F5100 & 0.366 & 0.273 & 0.365 & 0.287 & 0.355 & 0.288 \\
\addlinespace[2pt]
1990 & Total & 0.613 & 0.490 & 0.594 & 0.453 & 0.605 & 0.518 \\
1990 & F5100 & 0.448 & 0.298 & 0.418 & 0.278 & 0.432 & 0.300 \\
\addlinespace[2pt]
1991 & Total & 0.481 & 0.337 & 0.458 & 0.356 & 0.557 & 0.354 \\
1991 & F5100 & 0.388 & 0.324 & 0.398 & 0.306 & 0.452 & 0.352 \\
\addlinespace[2pt]
1992 & Total & 0.651 & 0.491 & 0.625 & 0.463 & 0.695 & 0.569 \\
1992 & F5100 & 0.505 & 0.384 & 0.467 & 0.361 & 0.515 & 0.400 \\
\addlinespace[2pt]
1993 & Total & 0.702 & 0.513 & 0.696 & 0.465 & 0.703 & 0.543 \\
1993 & F5100 & 0.601 & 0.400 & 0.568 & 0.370 & 0.620 & 0.413 \\
\bottomrule
\end{tabular}
\end{SURDtable}

\subsection{Sampling and estimator scope}
\label{sec:limits}
The component rankings identify candidate influences in the supplied observational decomposition. Native spectroscopy supplies 32--60 dates per season; interpolation yields 118--261 autocorrelated tuples rather than that many independent measurements. Flux errors, sampling-matched null distributions, peak-lag uncertainties and multiple-scan significance have not been estimated. Changing the flux-state count also changes the histogram occupancy and the random variables being predicted. The reported cross-bin recurrence and configuration comparisons therefore characterize sensitivity rather than statistical confidence.

The repository's independent categorical controls quantify a relevant estimator effect: at 118 tuples, median empirical leakage is 0.902 with two states and 0.426 with three even though the true leakage is one; at 261 tuples the values are 0.958 and 0.666. The archived audit also finds component-allocation sensitivity to predictor ordering in some categorical controls, associated with floating-point comparisons; its impact on these AGN histograms remains unquantified. Conservation of the full information budget checks arithmetic but does not independently validate that allocation. The source package retains these diagnostics. Matched resampling, flux-error propagation and a predictor-permutation audit of the actual histograms are the appropriate next tests of the specific candidate influences identified here. Also, Alternative approaches, eg. nearest-neighbour mutual-information estimators \citep{Kraskov2004}, provide useful methodological comparisons, although they are not employed in the present analysis are worth comparing in future tests.

\section{Discussion and conclusions}
\label{sec:epoch_conclusions}
The seasonal SURD analysis identifies which individual histories and combinations carry information about each future continuum or line state. The principal results are the named components and their simultaneous leakage, resolved across both four-variable sets, both flux discretizations and the three lag configurations.
\begin{enumerate}
\item The 1989 continuum--core channel is the clearest unique influence: common-lag $U(F)\to C$ reaches $g=0.799/0.387$ with leakage $0.047/0.276$ for two/three states. Its two-state contribution reaches 0.872 under either wing offset. The continuum is the largest cross-variable unique source for 12 of the 15 epoch--line-target pairs under both state counts.
\item The 1992 red-wing history carries a distinguished unique contribution to the future blue wing in both subsets and both state counts, with common-lag peaks $g=0.106$--0.166. The reverse unique component is zero in these scans, and the red history also informs the future integrated line. The channel persists with the blue history shifted and weakens when the red history is moved five days further back, identifying its temporal dependence.
\item Individual synergistic influences identify informative predictor pairs even when each component accounts for a moderate fraction of the target entropy. Examples include $S(F,R)\to B$ in 1989, $S(F,B)$ and $S(F,C)$ for the 1990 core, and $S(F,R)\to C$ in 1993. For the 1991 core, $S(F,B)$ and $S(F,R)$ contribute 0.187 and 0.125 at 20~d with leakage 0.087 while $U(F)=0$. A change to red-leading histories increases the latter joint term to 0.252 at the same base lag with two states.
\item Strong shared components identify opportunities for simpler predictive models. In the 1992 blue target, common-lag $R(H,B,C,R)$ reaches $g=0.681/0.709$. Specific $R(B,C)$ and $R(C,R)$ contributions reach approximately 0.22--0.38 in the highlighted unequal-lag cases. Either member supplies the shared contribution; the remaining unique and joint terms identify what retaining both may contribute, while a matched predictor-omission test determines whether either can be dropped.
\item Leakage identifies the unresolved part alongside these pathways. The largest two-state value, 0.967, occurs for the 1992 total-set red target at base lag 30~d with red-leading histories, while the 1993 scans have the largest mean common-lag leakage. These targets motivate additional driver measurements and longer temporal histories. The small leakage accompanying the strong 1989 continuum--core influence and the 1991 joint core terms identifies complementary cases in which the observed histories capture much of the categorical variability.
\end{enumerate}
The resulting picture is a season-dependent set of possible causal influences: continuum-specific driving, asymmetric wing information, joint continuum--line contributions and shared response among line bands. The repeatable continuum-associated red-before-blue ordering in 1989 and 1992 connects this information structure to reverberation observations, while the named SURD components reveal how that ordering is expressed in individual or joint histories. This complements the BLR response and kinematic measurements of \citet{Lu2022} and \citet{Xi2025}. The component catalogue and leakage measurements provide concrete targets for testing the driving continuum, the evolving line response and reduced predictive descriptions of NGC~5548.

\section*{Data and code availability}
The historical continuum measurements are available through the AGN Watch compilation, and the corrected line fluxes are published by \citet{WandersPeterson1996}. The archived SURD scans, atlases and diagnostics are in \url{https://github.com/am610/AGN_SURD}, under \texttt{collaborator\_rebuild}. This analysis uses commit \href{https://github.com/am610/AGN_SURD/tree/895a826512cc2f866035cc5b7b4430e35f28ab44}{\texttt{895a826512cc}} (full identifier in the accompanying reproducibility manifest). The principal inputs are \texttt{epoch\_results/individual\_components.csv.gz} and \texttt{epoch\_results/support\_and\_leakage.csv}; control values come from \texttt{calibration\_review/summary.csv} and the implementation caveat from \texttt{ESTIMATOR\_CALIBRATION.md}. The accompanying source package includes the manuscript tables, figure data, extended summary CSVs and scripts used for the named-component rankings, directional comparisons and simultaneous leakage calculations. A supplementary catalogue gives the top three cross-unique, redundant and synergistic components for all unequal-lag cases; full-precision CSVs retain every component maximum under both normalizations and all self-history terms. These are recalculations of archived decompositions, not a fresh SURD fit to native light curves.

\section*{Acknowledgements}

AM acknowledges the computational resources provided by the
National Center for Supercomputing Applications (NCSA),
which were used to carry out the simulations presented in this work.

\bibliographystyle{mnras}
\bibliography{refs_integrated}

\clearpage
\onecolumn
\appendix
\section{Supplementary information budgets}
\label{app:budgets}
The principal analysis resolves named components. Figure~\ref{fig:budgets} retains the earlier aggregate view as a supplementary consistency check: panels (a,b) compare the continuum-unique $f$ and $g$ maxima for the 1989 core; panel (c) shows why zero unique continuum information in 1991 can coexist with shared and joint information; panel (d) displays the matched-set leakage and synergy-budget changes. The individual pathways responsible for these budgets are identified in Tables~\ref{tab:synergistic_total_wings}--\ref{tab:redundant_continuum_wings}. Replacing $H$ by $F$ lowers lag-averaged leakage in all 90 matched epoch--line-target--configuration--bin comparisons, with common-lag mean reductions of 0.205/0.160 for two/three states. Equal-count common-lag epochs 1989--1992 give reductions of 0.231/0.170. These fitted information differences complement the named-component interpretation.
\begin{minipage}{\textwidth}\makeatletter\let\@makecaption\SFB@makefigurecaption\makeatother
\centering
\includegraphics[width=\textwidth]{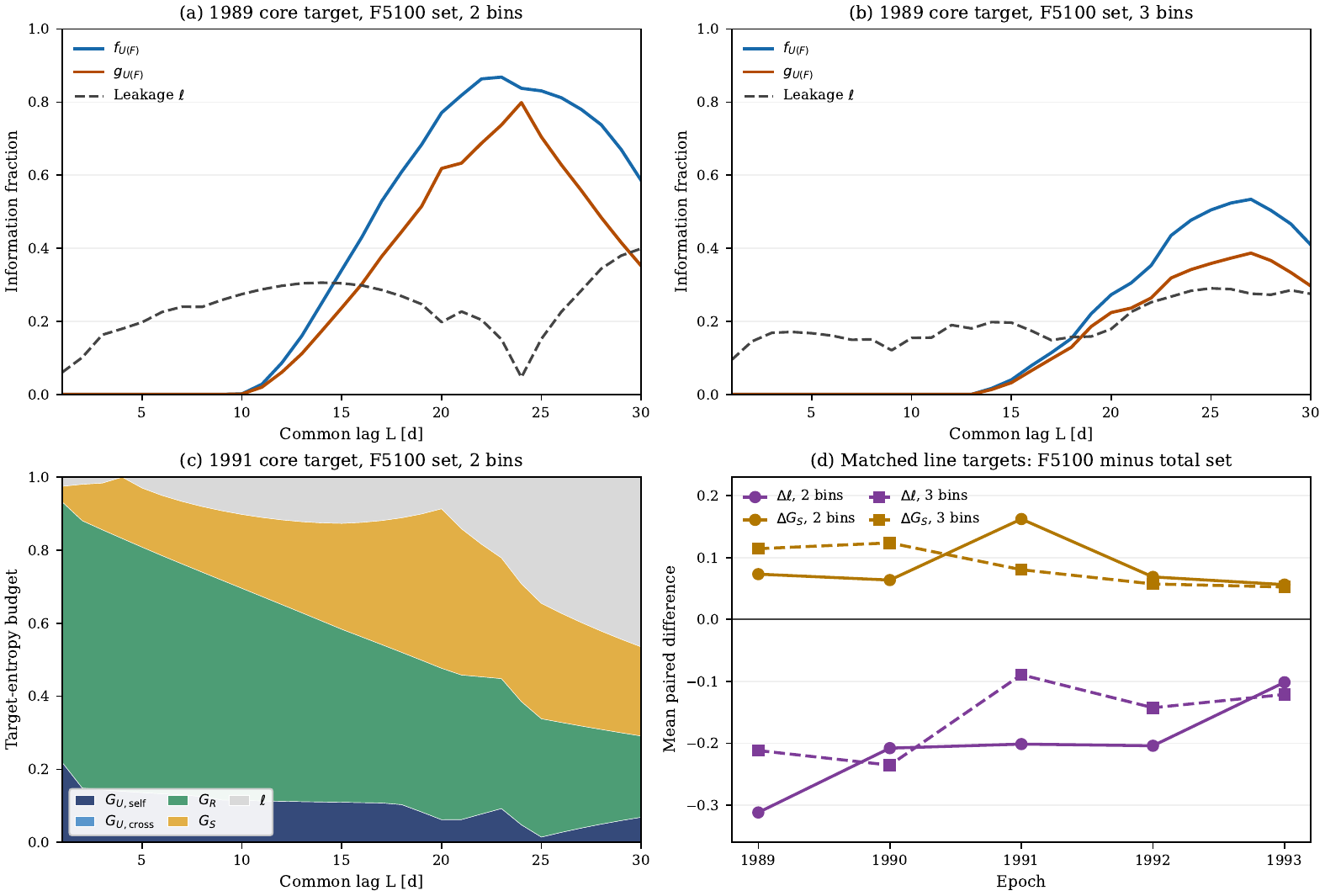}
\captionof{figure}{Information budgets calculated from the archived common-lag scans. Panels (a,b) show $f_{U(F)}$, $g_{U(F)}$ and simultaneous leakage for the 1989 core target in the F5100 set, with two and three bins. Panel (c) partitions the 1991 core's target entropy into unique self, unique cross-variable, redundant and synergistic budgets and leakage. Panel (d) compares the same three line targets between sets, averaging F5100-set minus total-set differences over targets and common lags. Lines guide the eye; there are no uncertainty estimates. The difference is not an intervention on the continuum.}
\label{fig:budgets}
\end{minipage}\clearpage
\onecolumn
\section{Dominant individual components for every common-lag target}
\label{app:dominant}
Table~\ref{tab:dominant} identifies, for every epoch, set, target and bin choice, the individual component with the largest $g_a$ anywhere in the common-lag scan. This criterion differs from choosing the largest $f_a$ or the largest sum over a component family. The corresponding leakage is evaluated at that component's maximizing lag. The table therefore supplies a consistent overview of each target without interpreting a large fraction of a small explained budget as a large target-entropy contribution.
\begin{minipage}{\textwidth}\makeatletter\let\@makecaption\SFB@maketablecaption\makeatother
\centering
\captionof{table}{Dominant individual common-lag component by maximum $g_a$ over all components and 1--30 observed days. Each bin block gives the component, maximizing lag, target-entropy fraction $g$ and simultaneous leakage $\ell$. $U$ denotes a unique component and $R$ a redundant group. H and F in the set column denote the total set and F5100 set, respectively; $C$ always denotes the line core. All entries are calculated from the archived CSV values.}
\label{tab:dominant}
\footnotesize
\begin{tabular}{ccclrrrlrrr}
\toprule
& & & \multicolumn{4}{c}{Two bins} & \multicolumn{4}{c}{Three bins} \\ \cmidrule(lr){4-7}\cmidrule(lr){8-11} Set & Epoch & Target & Component & $L$ & $g$ & $\ell$ & Component & $L$ & $g$ & $\ell$ \\
\midrule
H & 1989 & $H$ & $R(H,B,C,R)$ & 2 & 0.471 & 0.078 & $R(H,B,C,R)$ & 4 & 0.417 & 0.191 \\
H &  & $B$ & $R(H,B,C,R)$ & 7 & 0.416 & 0.230 & $R(H,B,C,R)$ & 6 & 0.388 & 0.226 \\
H &  & $C$ & $R(H,B,C,R)$ & 2 & 0.485 & 0.123 & $R(H,B,C,R)$ & 6 & 0.357 & 0.251 \\
H &  & $R$ & $R(H,B,C,R)$ & 2 & 0.298 & 0.250 & $U(R)$ & 1 & 0.394 & 0.185 \\
\addlinespace[2pt]
H & 1990 & $H$ & $R(H,B,C,R)$ & 1 & 0.619 & 0.059 & $R(H,B,C,R)$ & 2 & 0.633 & 0.085 \\
H &  & $B$ & $R(H,B,C,R)$ & 5 & 0.544 & 0.187 & $R(H,B,C,R)$ & 3 & 0.470 & 0.271 \\
H &  & $C$ & $R(H,B,C,R)$ & 1 & 0.492 & 0.095 & $R(H,B,C,R)$ & 3 & 0.545 & 0.123 \\
H &  & $R$ & $R(H,B,C,R)$ & 1 & 0.478 & 0.088 & $R(H,B,C,R)$ & 3 & 0.509 & 0.251 \\
\addlinespace[2pt]
H & 1991 & $H$ & $R(H,B,C,R)$ & 3 & 0.638 & 0.107 & $R(H,B,C)$ & 2 & 0.399 & 0.059 \\
H &  & $B$ & $R(H,B,C,R)$ & 4 & 0.627 & 0.101 & $R(H,B,C,R)$ & 6 & 0.375 & 0.145 \\
H &  & $C$ & $R(H,B,C,R)$ & 4 & 0.594 & 0.028 & $R(H,B,C)$ & 2 & 0.402 & 0.068 \\
H &  & $R$ & $R(H,B,C,R)$ & 1 & 0.442 & 0.142 & $U(R)$ & 1 & 0.432 & 0.169 \\
\addlinespace[2pt]
H & 1992 & $H$ & $R(H,B,C,R)$ & 1 & 0.679 & 0.042 & $R(H,B,C,R)$ & 1 & 0.612 & 0.073 \\
H &  & $B$ & $R(H,B,C,R)$ & 2 & 0.681 & 0.100 & $R(H,B,C,R)$ & 3 & 0.709 & 0.076 \\
H &  & $C$ & $R(H,B,C,R)$ & 1 & 0.611 & 0.083 & $R(H,B,C,R)$ & 1 & 0.635 & 0.044 \\
H &  & $R$ & $R(H,B,C,R)$ & 1 & 0.522 & 0.141 & $R(H,B,C,R)$ & 1 & 0.544 & 0.105 \\
\addlinespace[2pt]
H & 1993 & $H$ & $U(H)$ & 1 & 0.349 & 0.235 & $R(H,B,C,R)$ & 1 & 0.227 & 0.171 \\
H &  & $B$ & $U(B)$ & 1 & 0.296 & 0.278 & $R(H,B,C)$ & 2 & 0.309 & 0.272 \\
H &  & $C$ & $U(C)$ & 1 & 0.422 & 0.202 & $U(C)$ & 1 & 0.249 & 0.124 \\
H &  & $R$ & $U(R)$ & 1 & 0.575 & 0.233 & $U(R)$ & 1 & 0.427 & 0.292 \\
\midrule
F & 1989 & $F$ & $U(F)$ & 1 & 0.672 & 0.218 & $U(F)$ & 1 & 0.503 & 0.234 \\
F &  & $B$ & $U(F)$ & 26 & 0.455 & 0.370 & $U(F)$ & 27 & 0.365 & 0.233 \\
F &  & $C$ & $U(F)$ & 24 & 0.799 & 0.047 & $U(F)$ & 27 & 0.387 & 0.276 \\
F &  & $R$ & $U(F)$ & 22 & 0.598 & 0.248 & $U(R)$ & 1 & 0.406 & 0.157 \\
\addlinespace[2pt]
F & 1990 & $F$ & $U(F)$ & 1 & 0.637 & 0.224 & $U(F)$ & 1 & 0.639 & 0.191 \\
F &  & $B$ & $U(F)$ & 20 & 0.480 & 0.390 & $R(F,B,C,R)$ & 11 & 0.329 & 0.264 \\
F &  & $C$ & $R(B,C,R)$ & 1 & 0.408 & 0.079 & $R(B,C,R)$ & 1 & 0.454 & 0.027 \\
F &  & $R$ & $R(B,C,R)$ & 1 & 0.398 & 0.094 & $R(B,C,R)$ & 1 & 0.389 & 0.118 \\
\addlinespace[2pt]
F & 1991 & $F$ & $U(F)$ & 1 & 0.560 & 0.295 & $U(F)$ & 1 & 0.575 & 0.174 \\
F &  & $B$ & $R(B,C,R)$ & 4 & 0.515 & 0.079 & $R(B,C)$ & 1 & 0.346 & 0.030 \\
F &  & $C$ & $R(B,C,R)$ & 4 & 0.458 & 0.000 & $R(B,C)$ & 2 & 0.403 & 0.059 \\
F &  & $R$ & $U(R)$ & 1 & 0.373 & 0.120 & $U(R)$ & 1 & 0.458 & 0.146 \\
\addlinespace[2pt]
F & 1992 & $F$ & $U(F)$ & 1 & 0.360 & 0.277 & $R(F,B,C,R)$ & 1 & 0.319 & 0.200 \\
F &  & $B$ & $R(F,B,C,R)$ & 7 & 0.437 & 0.155 & $R(F,B,C,R)$ & 8 & 0.481 & 0.211 \\
F &  & $C$ & $R(F,B,C,R)$ & 4 & 0.428 & 0.214 & $R(F,B,C,R)$ & 3 & 0.454 & 0.143 \\
F &  & $R$ & $R(F,B,C,R)$ & 3 & 0.409 & 0.237 & $R(F,B,C,R)$ & 1 & 0.515 & 0.067 \\
\addlinespace[2pt]
F & 1993 & $F$ & $U(F)$ & 1 & 0.527 & 0.378 & $U(F)$ & 1 & 0.519 & 0.244 \\
F &  & $B$ & $U(B)$ & 1 & 0.396 & 0.285 & $U(B)$ & 1 & 0.326 & 0.161 \\
F &  & $C$ & $U(C)$ & 1 & 0.566 & 0.191 & $U(C)$ & 1 & 0.385 & 0.091 \\
F &  & $R$ & $U(R)$ & 1 & 0.667 & 0.238 & $U(R)$ & 1 & 0.473 & 0.264 \\
\bottomrule
\end{tabular}
\end{minipage}

\label{lastpage}
\end{document}